\RequirePackage{rotating}

\documentclass[11pt,a4paper,fleqn]{article}
\usepackage[utf8]{inputenc}
\usepackage{authblk}
\usepackage{geometry}
\usepackage{xcolor}
\usepackage[round]{natbib}
\usepackage{graphicx}
\usepackage{pdflscape}  
\usepackage{appendix}
\usepackage{threeparttable}
\usepackage{booktabs}
\usepackage{subcaption}

\usepackage{amssymb}
\usepackage{amsbsy}
\usepackage{bm}
\usepackage{amsmath}
\usepackage{amsthm}
\usepackage{graphicx}
\usepackage{mathtools}
\usepackage{setspace}
\usepackage{color, colortbl}
\definecolor{ashgrey}{rgb}{0.7, 0.75, 0.71}
\definecolor{columbiablue}{rgb}{0.61, 0.87, 1.0}
\definecolor{coral}{rgb}{1.0, 0.5, 0.31}

\definecolor{colBVAR}{HTML}{bababa}
\definecolor{colBART}{HTML}{d7191c}
\definecolor{colmixBART}{HTML}{fdae61}
\definecolor{colerrorBART}{HTML}{abd9e9}
\definecolor{colfullBART}{HTML}{2c7bb6}

\definecolor{colcons}{HTML}{e31a1c}
\definecolor{colSV}{HTML}{a6cee3}
\definecolor{colhBART}{HTML}{1f78b4}

\usepackage{enumitem}
\newlist{steps}{enumerate}{1}
\setlist[steps,1]{label = Step \arabic*:}

\usepackage{adjustbox}
\usepackage{dcolumn}
\newcolumntype{d}[1]{D..{#1}} % for alignment of numbers on decimal marker

\usepackage{caption}
\usepackage{subcaption}
\definecolor{nblue}{HTML}{000660}
\usepackage[colorlinks=true,urlcolor=nblue,linkcolor=nblue,citecolor=nblue]{hyperref}

\newcommand*{\myeqref}[2][Eq.~]{%
  \hyperref[{#2}]{#1(\ref*{#2})}%
}
\def\equationautorefname#1#2\null{%
  Eq.#1(#2\null)%
}

\begin{document}
\title{\textbf{Investigating Growth at Risk Using a Multi-country Non-parametric Quantile Factor Model}\thanks{
%\textit{Corresponding author}: Name. Affiliation. \textit{Address}: Address. \textit{Email}: \href{mailto:name@something.com}{name@something.com}. 
The views expressed herein are solely those of the authors and do not necessarily reflect the views of the Federal Reserve Bank of Cleveland or the Federal Reserve System. Marcellino thanks MIUR -- PRIN Bando 2017 -- prot.\ 2017TA7TYC  for financial support; Huber and Pfarrhofer gratefully acknowledge financial support from the Austrian Science Fund (FWF, grant no. ZK 35).}}
%We would like to thank .... for helpful comments.

\author[a]{Todd E. \textsc{Clark}}
\author[b]{Florian \textsc{Huber}}
\author[c]{Gary \textsc{Koop}}
\author[d]{Massimiliano \textsc{Marcellino}}
\author[b]{Michael \textsc{Pfarrhofer}}
\affil[a]{\textit{Federal Reserve Bank of Cleveland}}
\affil[b]{\textit{University of Salzburg}}
\affil[c]{\textit{University of Strathclyde}}
\affil[d]{\textit{Bocconi University, IGIER and CEPR}}
\date{\today}

\maketitle\thispagestyle{empty}\normalsize\vspace*{-2em}\small\linespread{1.5}
\begin{center}
\begin{minipage}{0.8\textwidth}
\noindent\small We develop a Bayesian non-parametric quantile panel regression model. Within each quantile, the response function is a convex combination of a linear model and a non-linear function, which we approximate using Bayesian Additive Regression Trees (BART). Cross-sectional information at the $p^{th}$ quantile is captured through a conditionally heteroscedastic latent factor.
 The non-parametric feature of our model enhances flexibility, while the panel feature, by exploiting cross-country information, increases the number of observations in the tails. We develop Bayesian Markov chain Monte Carlo (MCMC) methods for estimation and forecasting with our quantile factor BART model (QF-BART), and apply them to study growth at risk dynamics in a panel of 11 advanced economies. 

\end{minipage}
\end{center}

\normalsize\newpage
\section{Introduction}
Empirical macroeconomics has seen an upsurge of interest in modeling the tails of predictive distributions. A recent influential paper is \citet[ABG,][]{adrian2019vulnerable}, hereafter ABG, which investigated the impact of financial conditions on the conditional distribution of GDP growth and found it to be important in the lower quantiles. Both prior and subsequent to ABG, a large literature has emerged using quantile regression methods  to forecast tail risks to economic growth (see, among many others, \citet{adrian2018term, cook2019assessing, denicolo2017forecasting, Ferrara2019, giglio2016systemic, gonzalez2019growth, delle2020modeling, plagborg2020growth, reichlin2020financial, figueres2020vulnerable}; and \citet{mazzi2019nowcasting}). Other studies consider tail risks to other macroeconomic variables such as unemployment or inflation (e.g. \citet{galbraith2019asymmetry}, \citet{kiley2018unemployment}, \citet{ghysels2018quantile}, \citet{manzan2015forecasting}, \citet{gaglianone2012constructing}, \citet{korobilis2017quantile}, \citet{manzan2013macroeconomic,manzan2015asymmetric}, \citet{korobilis2021} and \citet{pfarrhofer2021}).

The existing literature, with few exceptions, uses quantile models for a single variable of interest. These models are specified conditional on a specific quantile and assume a linear relationship between the predictors and the quantile function of some outcome variable.\footnote{Examples of exceptions include \citet{korobilis2021} and \citet{pfarrhofer2021}, which assume time variation in the quantile regression coefficients. However, even these papers are single-equation and assume particular parametric forms for the time-variation. A multiple-equation exception is \citet{adrian2018term}, which exploits information in the term structure for an empirical application involving linear panel quantile regression.}   For macroeconomic data this assumption might be warranted in normal times but in turbulent times it could be that regression relationships change or turn non-linear. Moreover, often several variables rather than a single one are of interest, and in these cases a joint model would be preferable. These observations motivate the model we develop in the present paper. 

In contrast to much of the existing literature we propose a non-parametric model which involves multiple equations and allows for assessing whether the quantile response function is linear or unknown and possibly highly non-linear. In particular, the model we propose is a multi-country, non-parametric quantile regression, which we then use to  investigate growth at risk in a panel of 11 advanced economies. 

The justification for adopting non-parametric methods is provided by \citet{huber2020nowcasting} and \citet{chkmp2021tail}, which found Bayesian non-parametric vector autoregressions (VARs) to be able to successfully model the tails of predictive densities of macroeconomic variables in a flexible and accurate manner. These papers  found that Bayesian Additive Regression Trees (BART) are an effective non-parametric method that is particularly useful in crisis times (e.g., the Financial Crisis or the Covid-19 pandemic) when growth at risk issues are of particular importance. However, in normal periods, the predictive gains from using BART are more muted (and sometimes negative). In the present paper we extend the BART regression methods used in these papers to the quantile BART case.  Since the predictive gains of BART vary over the business cycle, we assume that within each quantile, the response function is a convex combination of a linear model and some unknown non-linear function, which we approximate using BART. Studies such as \citet{TaddyKottas2010JBES} have developed other Bayesian approaches to nonparametric model-based quantile regression.

The justification for use of a multi-country model is that a panel dimension can often improve forecasts with respect to single country models; see, among many others, \citet{BaiCarrieroClarkMarcellino2020WP} and \citet{fhkp2021}. Moreover, and specifically for the quantile case, macroeconomic data sets are short, leading to a small number of observations in the tails of the distribution. We develop a model for the $p^{th}$ quantile that includes a factor that summarizes the available cross-country information at that quantile. In addition, as indicated below, our Bayesian model specification has features that allow information from other countries to inform estimates for a given country. Exploiting this cross-country information through a pooling prior improves predictive accuracy  by parsimoniously including international information to inform coefficients associated with domestic quantities.

We then develop Bayesian Markov Chain Monte Carlo (MCMC) methods for estimation and forecasting with our quantile factor BART model (QF-BART). These methods are scalable to large panels with a potentially large number of exogenous regressors. 

In terms of empirical results, our proposed models commonly improve on the benchmark single country linear quantile model in recursive growth forecast comparisons, more so in the tails than near the center of the distribution.  Importantly, estimating the weight assigned to the BART component of the model as compared to the linear component is helpful to forecast accuracy. The estimated combination weight is smaller (i.e., the model is more linear) for the 25 and 75 quantiles than in the tails, i.e., for the 5 and 95 quantiles. Moreover, some form of international information definitely pays off (either via the new pooling prior, or by outright including non-domestic series). The effects of the common (volatility) factor are also relevant, as it seems to explain a large fraction of the forecast error variance in most countries, in particular in the tails. A shock to this factor, which can be interpreted as an uncertainty shock, has different (stronger negative) effects in the left tail than the right tail of the growth distribution.   In the left tail, the BART piece with an estimated weight tends to slightly mitigate the effects of the shock. Finally, a financial shock in the US spills over to other countries. There is asymmetry in the responses in the sense that a positive shock affects the growth quantiles, whereas a negative shock’s effects are not as sharp.  Moreover, the effects and asymmetry are more pronounced in the 2010-19 period than earlier.

The remainder of the paper is structured as follows. The next section defines and motivates our QF-BART model including its prior and discusses MCMC estimation. The third section contains our empirical work while the fourth section concludes the paper.

\section{A multi-country non-parametric regression model}\label{sec: econometrics}
We model the joint distribution of (for simplicity, demeaned) output growth for a panel of $N$ countries. These are stored in an $N-$dimensional vector $\bm y_t = ( y'_{1t}, \dots,  y'_{Nt})'$ with $ y_{it}$ denoting time $t$ output growth in country $i$. Domestic real activity might depend on the lags of $\bm y_{t}$ as well as other, exogenous factors.  These pre-determined quantities are included in a $K$-dimensional vector $\bm x_{it}$. We adopt a notational convention where $\bm x_{it}$ is structured such that $J$ domestic quantities for country $i$ are always ordered first, followed by all $K-J$ non-domestic variables. We assume that $y_{it}$ follows a  quantile regression model which, for the $p^{th}$ quantile, is given by:
\begin{equation}
    y_{it} = \omega_{ip} g_{ip}(\bm{x}_{it}) + (1-\omega_{ip})\bm{\beta}'_{ip} \bm{x}_{it} + {\lambda}_{ip} f_{pt} +  \epsilon_{ip,t}, \quad \epsilon_{ip,t}\sim\text{ALD}_p(\sigma_{ip}), \ i=1,\ldots,N, \label{eq: mixBART}
\end{equation}
with $g_{ip}: \mathbb{R}^K\rightarrow\mathbb{R} $ denoting unknown country-specific functions and $\bm \beta_{ip}$ is a $K \times 1$-dimensional vector of regression coefficients.
$\omega_{ip}$ is a  quantile and country-specific parameter that controls how much weight is placed on the non-linear part of the model. The case $\omega_{ip}=1$ would correspond to a fully non-linear model whereas $\omega_{ip}=0$ would be a (conditionally) linear  quantile regression specification. Contemporaneous relations across the elements in $\bm y_t$ are introduced through a static factor model with ${\lambda}_{ip}$ denoting the country-specific factor loading and $f_{pt}$ the corresponding international factor. Finally, $ \epsilon_{ip, t}$ follows an asymmetric Laplace distribution (ALD) scaled by a parameter $\sigma_{ip}$ with its $p^{th}$ quantile being equal to zero.\footnote{The density of the ALD$_p(\sigma_{p})$ is given by $p(1-p)/\sigma_p\exp(-\rho_p(\epsilon_t)/\sigma_p)$, where $\rho_p(x) = x(p-\mathbb{I}(x<0))$ is the check/loss function and $\mathbb{I}(\bullet)$ the usual indicator function. For details on the correspondence between the Bayesian and classical approach to inference in quantile regression, see \citet{yu2001bayesian}.}

We assume that the latent factor is uncorrelated over time and arises from a Gaussian distribution:
\begin{equation*}
    f_{pt} \sim \mathcal{N}(0, e^{h_{pt}}),
\end{equation*}
with $h_{pt}$ being a (logarithmic) variance that evolves according to an AR(1) process:
\begin{equation*}
    h_{pt} = \mu_p + \rho_p (h_{pt-1} - \mu_p) + \varsigma_p u_t, \quad u_t \sim \mathcal{N}(0, 1).
\end{equation*}
Here we let $\mu_p$ denote the unconditional mean, $\rho_p$ the autoregressive parameter, and $\varsigma^2_p$ the error variance of the log-volatility process.

This model possesses several features which should  improve not only its predictive capabilities but also allow for additional inferential opportunities. First, the presence of the quantile-specific weights allows for data-based selection of the degree of non-linearities across quantiles. Recent literature indicates that, in the lower tails of the distribution of output growth, macroeconomic relations change and might be subject to substantial non-linearities. While such non-linearities may be important in the extremes of a distribution,  linear models might describe the behavior well in tranquil periods of the business cycle (e.g., in the center of the distribution). Our model allows for this by setting the corresponding weights  $\omega_{ip}$ appropriately. Second, our model allows for lagged relations across countries. The key point to notice is that these dynamic interdependencies can differ  across quantiles. For instance, it could be that in the presence of a global adverse economic shock, cross-country dependencies are more important than in tranquil times. Third, the presence of a common static factor that exhibits conditional heteroscedasticity can capture contemporaneous relations across the elements in $\bm y_t$ for a specific quantile.  Moreover, since the factor is conditionally heteroscedastic, it can also control for sudden shifts in the conditional variance of the dependent variables. Inclusion of this stochastic volatility factor allows us to control for unobserved heterogeneity, a feature which might be extremely important during periods such as the recent Covid-19 pandemic (see, e.g., the discussion in \citet{carriero2021addressing}).

The model in \autoref{eq: mixBART} is quite general and nests several commonly used alternatives in the literature. For instance, setting $ \omega_{ip}=  0,  \lambda_{ip}=  0$ for all $p$, and setting $\bm x_{it}$ such that it includes only the first lag of GDP plus a (lagged) measure of financial conditions yields a model very closely related to the one proposed in \cite{adrian2019vulnerable}. Notice that this model essentially rules out cross-country relations. Setting $ \omega_{ip}=  1$ for all $p$ yields a non-parametric quantile regression model which, depending on $\lambda_{ip}$ and the choice of $\bm x_{it}$, allows for cross-country relations in a flexible manner. Setting $ \omega_{ip}=  0$ returns the linear quantile regression model but with a common volatility factor. 

\subsection{Approximating the unknown functions using BART}
We treat the function $g_{ip}$ as unknown and approximate it using BART \citep{chipman2010bart}.  Though other alternatives are possible, BART has been successfully employed in economics for forecasting financial time series in \cite{huber2020inference}, nowcasting GDP in selected European economies in \cite{huber2020nowcasting}, and tail forecasting of output, inflation, and unemployment in \citet{clark2021tail}. BART is a sum-of-trees model that approximates  $g_{ip}$ by summing over many individual trees that all take a simple form and act as ``weak learners.''  The BART approximation for $g_{ip}$ is given by:
\begin{equation*}
    g_{i p} \approx \hat{g}_{i p} = \sum_{s=1}^S v(\bm x_{it}|\mathcal{T}_{ip}^s, \bm \mu_{ip}^s),
\end{equation*}
with $v$ denoting a tree function that is determined by a tree structure $\mathcal{T}_{ip}^s$ and a vector of terminal node parameters $\bm \mu_{i p}^s$. This terminal node parameter vector has dimension $b_{ip}^s$.

The tree structure consists of multiple decision rules that ask whether a covariate exceeds a threshold and, according to these simple binary rules, produces (disjoint) partitions of the input space. These take the form $x_{ij, t} > c$ or $x_{ij,t} \le c$, with $x_{ij,t}$ denoting the $j^{th}$ element of $x_{it}$ and $c$ being a splitting/threshold value.  Sequences of these decision rules lead to a terminal node coupled with a corresponding terminal node parameter in $\bm \mu_{i p}^s$. 

When $S$ is large, the BART approximation is prone to overfitting if no further regularization is introduced. \cite{chipman2010bart} use regularization priors to force the trees $v$ to be simple. We achieve this through shrinkage priors on the tree structure and the terminal node parameters. Following \cite{chipman1998bayesian}, the prior on $\mathcal{T}_{ip}^s$ is obtained by constructing a tree-generating stochastic process. The prior $p(\mathcal{T}_{ip}^s)$ comprises of three aspects. First, tree complexity ultimately depends on the depth of intermediate nodes $d$. If $d$ is large, the tree is complex and thus might overfit the data. To force the individual trees to be simple, we assume that a given node at depth $d$ is non-terminal with probability proportional to:
\begin{equation*}
    \alpha~ (1+d)^{-\zeta}, 
\end{equation*}
where $\alpha$ is between $0$ and $1$ and $\zeta > 0$. Notice that this probability decreases in $d$: growing more complicated trees becomes unlikely if $d$ is already large. The amount of shrinkage is controlled by $\alpha$ and $\zeta$. These hyperparameters are often set to $\alpha=0.95$ and $\zeta = 2$, implying that trees with two or three terminal nodes receive over 80\% of total prior probability. \cite{chipman2010bart} found that, for over $40$ data sets, this choice performs well, and extensive cross-validation for $\alpha$ and $\zeta$ only improves predictive accuracy by small margins. The second and third aspect are concerned with how decision rules are constructed. To this end, we use discrete uniformly distributed priors to select the variables showing up in the decision rule as well as a uniform prior over the splitting/threshold values.

The second source of shrinkage is a Gaussian shrinkage prior on $\mu_{ij, p}^s$, the $j^{th}$ element of $\bm \mu^s_{ip}$. \cite{chipman2010bart} recommend scaling the prior using the range of the data. More specifically, let $y_{i, \text{min}}$ and $y_{i, \text{max}}$ denote the minimum and maximum of the observed data in country $i$. The corresponding Gaussian prior is then given by:
\begin{equation*}
    \mu_{ij, p}^s \sim \mathcal{N}\left(0, v^2_{ip}\right), \quad v_{ip} = \frac{y_{i, \text{max}} - y_{i, \text{min}}}{2 \gamma \sqrt{S}},
\end{equation*}
with $\gamma > 0$ being a prior scaling parameter, typically set equal to $2$. The prior implies that if the number of trees $S$ is large, the prior variance decreases and the amount explained by a single tree is decreased. This is consistent with the notion that each tree acts as a ``weak learner,'' explaining only a small share of variation in the response variable, but the ensemble model provides sufficient flexibility to capture even complicated conditional mean relations. Another feature, noted by \cite{huber2020nowcasting}, is that the prior variance increases in the range of the data. Hence, if outliers arise, the prior becomes increasingly loose and allows for more flexibility in terms of capturing observations far outside the range of past data.

The priors on the tree structures and the terminal node parameters constitute the main ingredients of BART. Since our model also features a linear part and additional coefficients, we also need to specify priors on $\bm \beta_{ip}$ and $\omega_{ip}$. We discuss them in the next sub-section.

\subsection{Priors on the remaining coefficients of the model}
On the coefficients $\bm \beta_{ip}$ we use a horseshoe-type prior on each element $\beta_{ij,p}$:
    \begin{equation*}
        \beta_{ij,p} \sim \mathcal{N}\left(\underline{\beta}_{jp},\varphi^2\psi_{ij, p}^2\right), \quad \varphi\sim\mathcal{C}^{+}(0,1), \quad \psi_{ij, p}\sim\mathcal{C}^{+}(0,1),
    \end{equation*}
where $\mathcal{C}^+$ denotes a half-Cauchy distribution,  $\psi_{ij,p}$ is a coefficient and quantile-specific scaling parameter, and $\varphi$ is a global shrinkage parameter that is common to all coefficients. Notice that the presence of $\varphi$ introduces dependencies across coefficients (including across countries) and across quantiles.  The key advantage is that the presence of the local shrinkage parameters $\psi_{ij, p}$ allows the detection of signals (i.e., non-zero or heterogeneous $\beta_{ij, p}$ over the cross section) even if $\varphi$ is close to zero.

The prior mean $\underline{\beta}_{jp}$ pools information over the cross section.  In our hierarchical specification, it is estimated from the data using a Gaussian prior for the domestic variables, and deterministically set to zero for non-domestic quantities:
\begin{equation*}
    \underline{\beta}_{jp} \sim\mathcal{N}(0,\tilde\varphi_j) \quad \text{for } j=1,\hdots,J, \qquad
    \underline{\beta}_{jp} = 0 \quad \text{for } j=J+1,\hdots,K.
\end{equation*}
The parameter $\tilde\varphi_j$ is the prior variance of the common mean, which we set to a weakly informative value of $10$ for the empirical application. We refer to this prior as the pooled horseshoe (HSP), while setting the common mean to a zero vector of size $K$ yields the conventional horseshoe (HS) that we consider as an alternative.

For the factor loadings $\lambda_{ip}$, we use a set of independent Gaussian priors for all $i, p$:
\begin{equation*}
    \lambda_{ip} \sim \mathcal{N}(0, 1).
\end{equation*}
Note that $\lambda_{ip}$ is a scalar and, hence, we  use this relatively non-informative prior rather than a prior such as the HS which is used to avoid over-parameterization as might occur with high dimensional parameters.

On the weights $\omega_{ip}$ we consider a Uniform prior:
\begin{equation}
    \omega_{ip} \sim \mathcal{U}(0, 1).
\end{equation}
This prior introduces no particular prior information on the amount of non-linearities. If we wish to be informative on $\omega_{ip}$ we can also use a Beta prior and specify the hyperparameters appropriately.

%This prior allows us to be informative on the amount of non-linearities by pushing $\omega_{ip}$ towards $1$ (i.e., a flexible conditional mean model) or $0$ (i.e., a linear quantile regression).

% Since we would like to let the data speak on how much weight should be placed on the BART model, in our empirical work we set  $a_\omega = b_\omega = 1$, leading to a uniform prior on the weights.

The remaining coefficients of the model relate to the error term. \citet{kozumi2011gibbs} write the ALD using a scale-location mixture of Gaussians:
\begin{equation*}
    \epsilon_{ip, t} = \theta_{p} \nu_{ip, t} + \tau_{p}\sqrt{\sigma_{ip} \nu_{ip,t}} e_{ip, t},
\end{equation*}
with  $\theta_{p} = \frac{1-2p}{p (1-p)},\quad \tau^2_{p} = \frac{2}{p(1-p)},\quad  \nu_{ip, t} = \sigma_{ip} z_{ip, t},\quad e_{ip,t} \sim \mathcal{N}(0, 1)$, and $z_{ip, t} \sim \text{Exp}(1)$. On the scale parameter $\sigma_{ip}$ we use an inverse Gamma prior:
\begin{equation*}
    \sigma_{ip} \sim \mathcal{G}^{-1}\left(\frac{a_\sigma}{2},\frac{b_\sigma}{2}\right),
\end{equation*}
with the relatively non-informative choices of $a_\sigma = 1$ and $b_\sigma = 1$. 

This completes the prior setup. In the next sub-section we briefly discuss the Markov chain Monte Carlo (MCMC) algorithm used to carry out estimation and inference.

\subsection{Full conditional posterior simulation}
We use Markov Chain Monte Carlo (MCMC) techniques to obtain a draw from the joint posterior of the latent quantities and coefficients of the model.  Specifically, the following steps of the algorithm are carried out for each equation (i.e., country) $i$ and quantile $p$:
\begin{itemize}
    \item \textbf{Sampling from $p(\mathcal{T}^s_{ip}| \bullet)$ and $p(\bm \mu^s_{ip}| \bullet)$}. The full conditional posterior of the tree structures takes no well-known form. \cite{chipman2010bart} propose a Bayesian backfitting strategy to set up a Metropolis Hastings (MH) algorithm to sample the trees individually, conditionally on the other $S-1$ trees. This step is carried out marginally of $\bm \mu^s_{ip}$. The terminal node parameters can then, under our conjugate prior, be simulated from a set of independent Gaussian distributions that take a well-known form.
    
    \item \textbf{Sampling $p(\bm \beta_{ip} | \bullet)$}. The regression coefficients are, conditional on the remaining parameters and latent states, simulated from a multivariate Gaussian posterior distribution with known moments:
    \begin{equation*}
        \bm \beta_{ip}|\bullet \sim \mathcal{N}(\overline{\bm \beta}_{ip}, \overline{\bm V}_{ip}), \quad \overline{\bm V}_{ip} = \left(\tilde{\bm X}'_{ip} \tilde{\bm X}_{ip} + \underline{\bm V}_{ip}^{-1}\right)^{-1}, \quad \overline{\bm \beta}_{ip} = \overline{\bm V}_{ip} \left(\underline{\bm V}_{ip}^{-1} \underline{\bm{\beta}}_{p} + \tilde{\bm X}'_{ip} \tilde{\bm y}_{ip}\right).
    \end{equation*}
    $\tilde{\bm y}_{ip}$ is a $T-$dimensional response vector with $t^{th}$ element given by $\tilde{y}_{ip, t} = (y_{it} - \omega_{ip} \hat{g}_{ip}(\bm x_{it}) - \lambda_{ip} f_{pt} - \theta_{p} \nu_{ip,t})/(\tau_{p} \sqrt{\sigma_{ip} \nu_{ip,t}})$, $\tilde{\bm X}_{ip}$ is a $T \times K$ matrix with typical row $\tilde{\bm x}_{ip, t} = ((1-\omega_{ip}) \bm x_{it})/(\tau_{p} \sqrt{\sigma_{ip} \nu_{ip,t}})$, and $\underline{\bm V}_{ip}$ is a prior variance matrix with main diagonal element $v_{ij,p}=\varphi^2 \psi^2_{ij,p}$. The prior mean $\underline{\bm{\beta}}_{p}=(\underline{\beta}_{1p},\hdots,\underline{\beta}_{Jp},\bm{0}'_{K-J})'$ collects the estimated common means $\underline{\beta}_{jp}$ in the corresponding position of the $J$ domestic variables with the remaining elements being zero for the HSP prior, while $\underline{\bm{\beta}}_{p}=\bm{0}_{K}$ for the conventional HS prior.
    
    \item \textbf{Sampling from} $p(\underline{\beta}_{jp}|\bullet)$. The posterior distribution for the non-zero elements for $j=1,\hdots,J$ of the prior mean for HSP is $\underline{\beta}_{jp}\sim\mathcal{N}\left(\overline{b}_{jp},\overline{v}_{j p}\right)$, with moments
    \begin{equation*}
        \overline{v}_{jp} = \left(\left(\sum_{i=1}^N \frac{1}{v_{ij,p}}\right) + \frac{1}{\varphi_j}\right)^{-1}, \quad \overline{b}_{jp} = \overline{v}_{j,p}\left(\sum_{i=1}^N \frac{\beta_{ij,p}}{v_{ij,p}}\right).
    \end{equation*}

\item \textbf{Sampling from $p(\omega_{ip} | \bullet)$}: The full conditional posterior of $\omega_{ip}$ takes no well-known form. Since the support of $\omega_{ip}$ is bounded and the target density univariate, we adopt a slice sampler (see, e.g., \cite{slicesampling}) that is straightforward to implement and mixes well. 

\item \textbf{Sampling from $p(\lambda_{ip} | \bullet)$.} The factor loadings are obtained by simulating from univariate Gaussian conditional posterior distributions:
\begin{equation*}
    \lambda_{ip}|\bullet \sim \mathcal{N}( \overline{\lambda}_{ip}, \overline{l}_{ip}), \quad \overline{l}_{ip} = (\hat{\bm f}'_p \hat{\bm f}_p + 1)^{-1}, \quad \overline{\lambda}_{ip}= \overline{l}_{ip} \hat{\bm f}'_p \hat{\bm y}_{ip}.
\end{equation*}
$\hat{\bm y}_{ip}$ denotes the $T \times 1$ response vector with typical $t^{th}$ element given by $\hat{y}_{ip,t} = (y_{it} - \omega_{ip} \hat{g}_{ip}(\bm x_{it}) - (1-\omega_{ip}) \bm \beta'_{ip} \bm x_{it})/(\tau_{p} \sqrt{\sigma_{ip} \nu_{ip,t}})$, and $\hat{\bm f}_p$ has typical element $\hat{f}_{pt}= f_{pt}/(\tau_{p} \sqrt{\sigma_{ip} \nu_{ip,t}})$. 

\item \textbf{Sampling from $p(\sigma_{ip}|\bullet)$.} \citet{kozumi2011gibbs} show that the conditional posterior of the scaling parameter $\sigma_{ip}$ follows an inverse Gamma distribution:
\begin{equation*}
    \sigma_{ip}|\bullet \sim \mathcal{G}^{-1}\left(\frac{\tilde{a}_{ip}}{2}, \frac{\tilde{b}_{ip}}{2}\right),
\end{equation*}
with $\tilde{a}_{ip} = a_\sigma + 3 T$, $\tilde{b}_{ip} = b_\sigma + 2\sum_{t=1}^T \nu_{ip, t} + \sum_{t=1}^T (w_{it} - \theta_{p} \nu_{ip,t})^2/(\tau^2_{p} \nu_{ip,t})$, and $w_{it} = y_{it} - \omega_{ip} \hat{g}_{ip}(\bm x_{it}) - (1-\omega_{ip}) \bm \beta'_{ip} x_{it} - \lambda_{ip} f_{pt}$.

%(y_{it} - \omega_{ip} \hat{g}_{ip}(\bm x_{it}) - (1-\omega_{ip}) \bm \beta'_{ip} x_{it} - \lambda_{ip} f_{pt} - \theta_{p} \nu_{ip,t})^2/$.

\item \textbf{Sampling from $p(\nu_{ip, t}|\bullet)$}. For each $t$, we simulate $\nu_{ip, t}$ from a generalized inverse Gaussian (GIG) posterior distribution:
\begin{equation*}
    \nu_{ip, t}|\bullet \sim \text{GIG}\left(\frac{1}{2}, \tilde{c}_{ip}, \tilde{d}_{ip}\right),
\end{equation*}
where $\tilde{c}_{ip} = w_{it}/(\tau^2_{p} \sigma_{ip})$ and $\tilde{d}_{ip} = 2/ \sigma_{ip} + \theta^2_{p}/(\tau^2_{p} \sigma_{ip})$.\footnote{The generalized inverse Gaussian (GIG) distribution is parameterized such that a random variable $X\sim\mathcal{GIG}(\lambda,\xi,\psi)$ has probability density function $f(x;\lambda,\xi,\psi)=x^{\lambda-1}\exp\left(-(\xi/x + \psi x)/2\right)$.}

\item \textbf{Sampling from $p(\psi^2_{ij,p}|\bullet)$}. The local, coefficient-specific scaling parameters are simulated using the scheme outlined in \citet{makalic2015simple}. Conditional on  auxiliary  shrinkage parameters $\xi$ and $\eta_{ij,p}$, the posterior of $\psi^2_{ij,p}$ is inverse Gamma distributed:
\begin{equation*}
    \psi^2_{ij,p} \sim \mathcal{G}^{-1}\left(1, \frac{1}{\eta_{ij,p}} + \frac{\left(\beta_{ij,p}-\underline{\beta}_{jp}\right)^2}{2 \varphi^2}\right), \quad  \eta_{ij,p}|\bullet \sim \mathcal{G}^{-1}\left(1, 1 + \frac{1}{\xi}\right).
\end{equation*}
\end{itemize}

These steps relate to the quantities we have to simulate for each country (or equation) and quantile. Next we turn to the quantities that we simulate per quantile and thus pool over countries.
\begin{itemize}
    \item \textbf{Sampling from $p(f_{pt}|\bullet)$}. For each $t$, we simulate $f_{pt}$ from a sequence of independent Gaussian posterior distributions as follows:
    \begin{equation*}
        f_{pt}|\bullet \sim \mathcal{N}(\bm A_{pt} \tilde{\bm y}_{pt}, e^{h_{pt}} - \bm A_{pt} \bm \Xi_{pt} \bm A'_{pt}), \quad \bm \Xi_{pt} = e^{h_{pt}} (\bm \lambda_p  \bm \lambda'_p) + \bm \Psi_{pt}, \quad \bm A_{pt} = e^{h_{pt}} \bm \lambda' \bm \Xi_{pt},
    \end{equation*}
    whereby $\bar{\bm y}_{pt}$ is a $N\times 1$ vector with  $\bar{y}_{ip,t} = y_{it} - \omega_{ip} \hat{g}_{ip}( \bm x_{it}) + (1-\omega_{ip}) \bm \beta'_{ip} \bm x_{it} - \theta_{p} \nu_{ip,t}$, $\bm \lambda_p =(\lambda_{1p}, \dots, \lambda_{Np})'$, and  $\bm \Psi_{pt} = \text{diag}(\tau^2_{p} \sigma_{1p} \nu_{1p, t}, \dots, \tau^2_{p} \sigma_{Np} \nu_{Np, t})$.
    \item \textbf{Sampling from $p(\bm h_{p}|\bullet)$ and $p(\mu_p, \rho_p, \varsigma_p|\bullet)$}. We sample the full history of log-volatilities $\bm h_p = (h_{p1}, \dots, h_{pT})'$ and the parameters of the state evolution equation using the efficient sampler proposed in \cite{kastner2014ancillarity}. This algorithm samples the log-volatilities, conditional on everything else, all without a loop.
\end{itemize}

The final step refers to the global shrinkage parameter of the horseshoe prior. This step pools information across all equations and quantiles. Since we rely on auxiliary random variables to obtain a well-known full conditional posterior distribution, we first simulate from $p(\xi| \bullet)$ and then from $p(\varphi|\bullet)$.
\begin{itemize}
    \item \textbf{Sampling from $p(\varphi^2 | \bullet)$ and $p(\xi|\bullet)$}. The conditional posteriors of the global shrinkage parameter $\varphi$ and the auxiliary global parameter $\xi$ are, respectively, inverse Gamma distributed:
    \begin{equation*}
        \varphi^2|\bullet \sim \mathcal{G}^{-1}\left(\frac{N K + 1}{2}, \frac{1}{\xi} + \frac{1}{2} \sum_{p} \sum_{i=1}^{N}\sum_{j=1}^K \frac{\left(\beta_{ij,p}-\underline{\beta}_{jp}\right)^2}{ \psi^2_{ij,p}}\right), \quad \xi|\bullet \sim \mathcal{G}^{-1}\left(1, 1 + \frac{1}{\varphi^2}\right).
    \end{equation*}
\end{itemize}
This completes our MCMC algorithm. In all our empirical work we repeat the different steps $30,000$ times and discard the first $15,000$ draws as burn-in. One key advantage of the present algorithm is that it is scalable to larger data sets (i.e., including more countries, additional endogenous variables, or more covariates) because, conditional on the factors and $\varphi$, the different posterior quantities are independent across equations and quantiles. 

\section{Empirical results}
In this section we first investigate whether our modeling approach improves upon a set of simpler, nested alternatives by means of a forecasting horse race. We then focus on international growth at risk dynamics in two ways. First, we analyze how GDP growth reacts to changes in the common factor. Afterwards, we focus on how a shock to US financial conditions spills over to the other economies in our sample.

%The next sub-section provides an overview of the data 

\subsection{Data overview, competing models and forecasting design}\label{subsec: data}
Our sample runs from 1975Q1 to 2020Q4. We use annualized quarterly growth rates of GDP data from the Main Economic Indicators (MEI) database, maintained by the OECD, and the composite indicator of systemic stress (CISS) by the European Central Bank (ECB). For data availability reasons we include Austria (AT), Denmark (DK), Finland (FI), France (FR), Germany (DE), Italy (IT), Netherlands (NL), Spain (ES), Sweden (SE), United Kingdom (UK), and the United States (US).

We estimate the models for $p\in\{0.05, 0.10,0.25,0.50,0.75,0.90, 095\}$. For each model we consider two different choices for the covariates. The first, which we label \texttt{CISS}, includes the CISS and a single lag of $y_{it}$ in $\bm x_{it}$, implying that $K=2$. The second includes cross-country information in $\bm x_{it}$ by including the first lag of GDP growth and the CISS of all countries; hence, $K=2N$. The latter is referred to as \texttt{CISS-CC} to indicate that the information set includes cross-country data. 

Since our model is quite flexible and nests several competing models, we also include a range of restricted variants of the general model outlined in Section \ref{sec: econometrics}.  First, we obtain the ABG model by using the CISS covariates and  setting $\bm \Lambda= \bm 0$ and  $\omega_{ip}=0$. We use frequentist methods to carry out estimation so as to be the same as ABG,  while we estimate all other models using Bayesian methods with either the HS or HSP prior (so, for example, we will consider both CISS-CC-HS and CISS-CC-HSP specifications). ABG will serve as our benchmark model to which we compare all other specifications. We then add features to this benchmark. We begin by remaining linear ($\omega_{ip}=0$) but adding the international factor to the ABG model by letting $\bm \Lambda \neq \bm 0$ in order to investigate whether it plays an empirically important role. All subsequent models also let $\bm \Lambda \neq \bm 0$.  We next investigate non-linearities by setting $\omega_{ip}=1$ and thus obtain a multi-country quantile BART model with a common international factor. Finally our most flexible model allows for $\omega_{ip}$ to be estimated from the data. An overview of all model specifications is provided in Table \ref{tab:models}.

We compute pseudo out-of-sample forecasts based on a holdout from 1990Q1 to 2020Q4 (so the initial training sample comprises $60$ quarters). We compute Quantile Scores (QS, for quantiles $0.1,0.25,0.5,0.75,0.9$) and quantile-weighted cumulative ranked probability scores (qw-CRPS, see \cite{gneiting2011comparing}) with five weighting schemes (``none'' refers to no weighting, i.e., conventional CRPS; both tails ``tails;'' left tail, ``left;'' right tail, ``right;'' and ``center''). We compute direct forecasts for $h\in\{1, 4\}$. 

\begin{table*}[ht]
\caption{Model overview.}\vspace*{-1.5em}
\begin{center}
\begin{small}
\begin{threeparttable}
\begin{tabular*}{\textwidth}{@{\extracolsep{\fill}} llll}
\toprule
\textbf{Data} & \textbf{Prior} & \textbf{Weights} & \textbf{Factor}\\
\midrule
\texttt{CISS} (domestic) & \texttt{HS} (shrinkage to zero) & $\omega=0$ (parametric) & $\bm{\Lambda}=0$ (independence) \\
\texttt{CISS-CC} (cross-country) & \texttt{HSP} (pooling cross-section) & $\omega=1$ (nonparametric) & $\bm{\Lambda}\neq0$ (dependence) \\
 &  & $\omega\in(0,1)$ (estimated) &  \\
\bottomrule
\end{tabular*}
\begin{tablenotes}[para,flushleft]
\scriptsize{\textit{Notes}: ``Data'' refers to the information set for individual country models. ``Prior'' indicates the prior on the parametric part of the model; we consider the conventional horseshoe prior (HS) shrinking towards zero and the pooled horseshoe (HSP) prior pushing the model towards cross-sectional homogeneity. ``Weights'' refers to the specification of the conditional quantile function: parametric, nonparametric, or whether we estimate weights on the parametric and nonparametric part. ``Factor'' indicates whether an international factor modeling the cross-sectional covariance structure within quantiles is present. We consider all possible combinations of these specification choices.}
\end{tablenotes}
\end{threeparttable}
\end{small}
\end{center}
\label{tab:models}
\end{table*}

\subsection{Tail forecasting results}\label{subsec: forecasting}
% ----------------------------------------------------------------------------------------------

Table \ref{fig:forecast_crps} reports the forecast comparison of the various models based on the relative qw-CRPS. Each cell in the heatmap shows the qw-CRPS relative to the ABG benchmark model. Numbers smaller than one indicate outperformance (green colored) vis-\'{a}-vis the ABG model whereas numbers exceeding one suggest a weaker performance (red colored) than the benchmark.

\begin{table}
    \includegraphics[width=\textwidth]{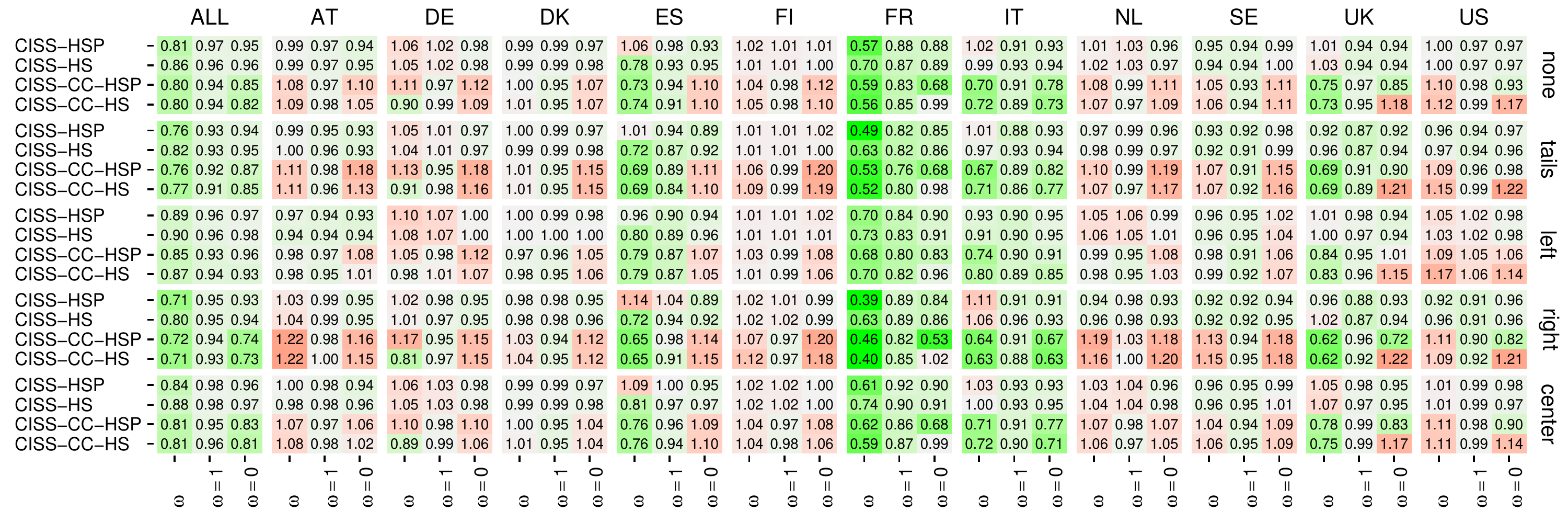}
    \caption{Relative quantile weighted cumulative ranked probability scores (CRPS) for $h=1$ and models with $\bm{\Lambda}\neq\bm{0}$. The results are benchmarked to \texttt{CISS} with $\bm{\Lambda}=\bm{0}$ and $\bm{\omega}=\bm{0}$. Lower ratios (shaded in green) indicate better performance (and vice versa, shaded in red).}\label{fig:forecast_crps}
\end{table}

Four main comments can be made. First, and focusing on aggregate results across countries, our proposed models commonly improve on the benchmark ABG model. The gains are about 20 percent when looking at the standard CRPS, decrease to about 10 to 15 percent in the left tail, and increase to about 30 percent in the right tail (based on additional results reported in the Appendix). Overall, these results suggest that at each quantile, and particularly in the right tail, it pays off to allow for non-linearities and for cross-country relations.

Second, while there are small differences between setting $\omega_{ip}=0$ (linear quantile) or $\omega_{ip}=1$ (BART quantile), there is often some benefit to estimating the weight $\omega_{ip}$, in turn allowing for both linear and BART pieces in the model. The key advantage of estimating $\omega_{ip}$ is that it combines the best of both worlds and thus translates into a model that is strongly non-linear and non-parametric in the tails and close to a linear quantile regression model in the center of the distribution. Such a behavior is beneficial if loss functions which evaluate the full predictive distribution are used.

Third, the HSP prior, that includes pooling, is typically better than HS, but the differences shrink or are eliminated once cross-country information is included in the model. It is noteworthy that once we use a pooling prior the predictive benefit of adding cross-country information directly diminishes sharply. This points towards the fact that, through pooling, our approach successfully picks up cross-sectional information in a very parsimonious manner.

Finally, there is some heterogeneity across the countries under analysis. In particular, for Spain, France, Italy, and the UK the results are broadly in line with those mentioned above. In contrast, for Austria and Sweden estimating the weight $\omega$ yields little gains (setting $\omega_{ip}=1$ is often best), and for the other countries it is overall difficult to beat the benchmark. 

\subsection{Estimated weights over time}
In the previous sub-section we have shown that our proposed framework yields forecast distributions which are often more precise than the ones obtained from the ABG benchmark and simpler nested alternatives. One key advantage of the model is that it allows for different weights $\omega_{ip}$ across countries and quantiles and this improves forecasts when the full predictive density is evaluated. In this sub-section, we investigate whether our intuition that non-linearities are relevant in the tails while linear models are adequate in the center of the distribution is supported by our model.

\begin{figure}[t]
    \includegraphics[width=\textwidth]{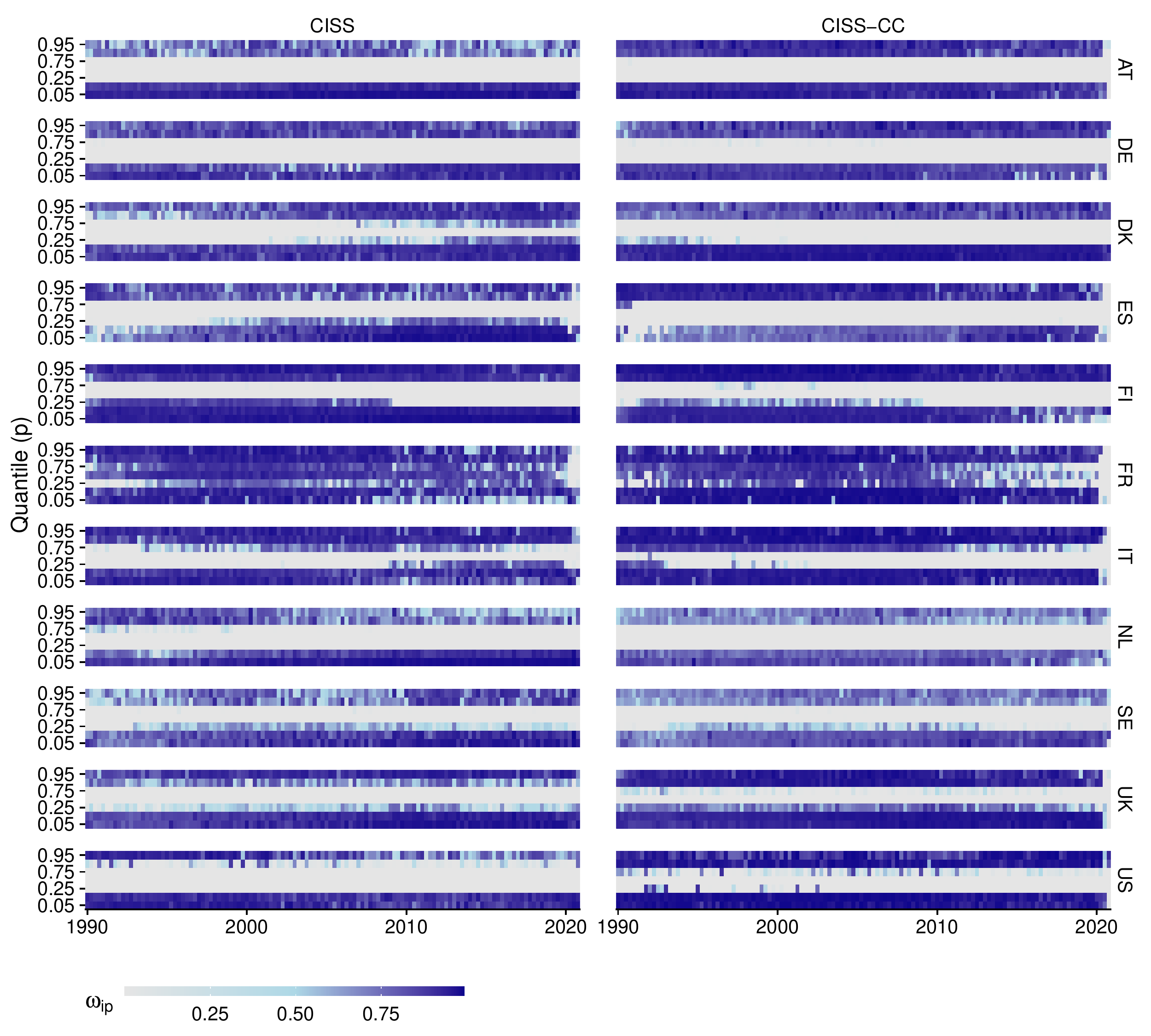}
    \caption{Non-parametric weights $\omega_{ip}$ for $\bm{\Lambda}\neq0$ with the HSP prior plotted over time with respect to the training/holdout samples ($h=1$).}\label{fig:modelweights}
\end{figure}

Figure \ref{fig:modelweights} reports the estimated weight $\omega_{ip}$ over our hold-out period. Darkblue cells indicate a weight close to one while gray shaded cells imply a weight close to zero. We focus on two models, the CISS and CISS-CC models coupled with the pooled Horseshoe prior (HSP).\footnote{The results for the remaining specifications look similar and may be found in the Appendix.}

It turns out that for most countries, our conjecture is confirmed. That is, we observe weights which approach unity if we move out in the tails (i.e., the model becomes more non-linear). When we focus on the center of the distribution, the combination weights approach zero (i.e., the model is linear). Comparing the right and left tails reveals that  the estimated weight is often larger for the 5 percent quantiles than the 95 percent quantiles. This indicates that non-linearites are important when our focus is on modeling sharp upswings in GDP growth but become even more important when interest centers on capturing downturns in GDP growth that are extreme (i.e., below the 5 percent quantile).

When we compare the model which does not utilize cross-country information (CISS) to the one which explicitly includes cross-sectional data (CISS-CC), we find only modest differences in combination weights. These differences are mainly related to somewhat smaller weights on the BART specification in the upper tail of the distribution (for some selected countries such as AT, DK, the UK, and the US), but the main finding that, in the center of the distribution, our model assigns no weight to the non-linear model  still holds.

Zooming into the different results reveals that most countries share the general dynamics described in the previous paragraphs (i.e., $\omega_{ip}$ close to one in the tails and $\omega_{ip} \approx 0$ in the center of the distribution). One exception to this broad-based finding is France, which displays larger combination weights across all quantiles. In addition, there exists temporal heterogeneity. For instance, in several countries we observe that, after the global financial crisis (and sometimes slightly earlier), combination weights decrease markedly in the upper tails of the distribution.

\subsection{The role of the common volatility factor}

In this sub-section, we investigate the common volatility factor across quantiles. In a first step, we assess the relevance of the  common factor volatility specification by considering time averages of variance decompositions. These are computed, by taking the Gaussian representation of the ALD (the distribution of $\epsilon_{ip,t}$ in Equation (\ref{eq: mixBART})), as follows:
\begin{equation*}
    \text{VD}_{ip, t} = \frac{\lambda^2_{ip} e^{h_{pt}}}{\lambda^2_{ip} e^{h_{pt}} + \mbox{Var}(\epsilon_{ip,t})},
\end{equation*}
with $\mbox{Var}(\epsilon_{ip,t})$ denoting the variance of $\epsilon_{ip,t}$. This decomposition provides information on the share of variation in the shocks (conditional on the quantile) that is explained through the common factor (similar to \citet{SW2005JEEA}).

Table \ref{tab: vardecomp} reports time averages of variance decompositions resulting from the CC-HSP model.  Interestingly, for most countries the commonality is larger and more substantial in the tails than at the center of the distribution, and a bit larger in the right than in the left tail. These larger contributions in extreme periods can be traced back to the fact that several of the recessions in our hold-out period can be viewed as shocks with a pronounced global dimension (such as the global financial crisis or the Covid-19 pandemic) and the factor is picking this up. 

Across countries, we find a considerable degree of homogeneity within country groups. For instance, Finland, Denmark, and Sweden feature commonalities that are very pronounced in the tails but decline once we approach the center of the distribution both from left and right.  The US and the UK share a rather similar pattern in terms of commonalities (high shares in the tails and for the median, smaller shares for the quantiles in between).

%In sum, the high explanatory power of the factor highlights the importance of this feature of the model in particular when modelling the tails.

\begin{table}[t]
\caption{Time averages of variance decompositions, $\omega$ sampled, CC model and HSP prior.} \label{tab: vardecomp}
\centering
\begin{tabular}{lrrrrrrrrrrrr}
 \toprule
Quantile ($p$) & ALL & AT & DE & DK & ES & FI & FR & IT & NL & SE & UK & US \\ 
  \midrule
0.05 & 0.83 & 0.85 & 0.84 & 0.79 & 0.83 & 0.77 & 0.86 & 0.84 & 0.84 & 0.86 & 0.87 & 0.82 \\ 
  0.1 & 0.73 & 0.77 & 0.72 & 0.63 & 0.77 & 0.61 & 0.80 & 0.76 & 0.73 & 0.76 & 0.77 & 0.75 \\ 
  0.25 & 0.55 & 0.63 & 0.59 & 0.42 & 0.61 & 0.38 & 0.56 & 0.61 & 0.56 & 0.53 & 0.61 & 0.58 \\ 
  0.5 & 0.67 & 0.83 & 0.68 & 0.30 & 0.95 & 0.21 & 0.98 & 0.90 & 0.56 & 0.40 & 0.89 & 0.71 \\ 
  0.75 & 0.70 & 0.79 & 0.76 & 0.50 & 0.77 & 0.48 & 0.80 & 0.81 & 0.69 & 0.59 & 0.78 & 0.69 \\ 
  0.9 & 0.79 & 0.83 & 0.85 & 0.68 & 0.81 & 0.65 & 0.85 & 0.84 & 0.79 & 0.75 & 0.82 & 0.77 \\ 
  0.95 & 0.89 & 0.91 & 0.91 & 0.82 & 0.90 & 0.82 & 0.92 & 0.91 & 0.89 & 0.87 & 0.92 & 0.87 \\ 
   \bottomrule
\end{tabular}
\end{table}

\begin{figure}[t]
    \centering
    \includegraphics[width=1.05\textwidth]{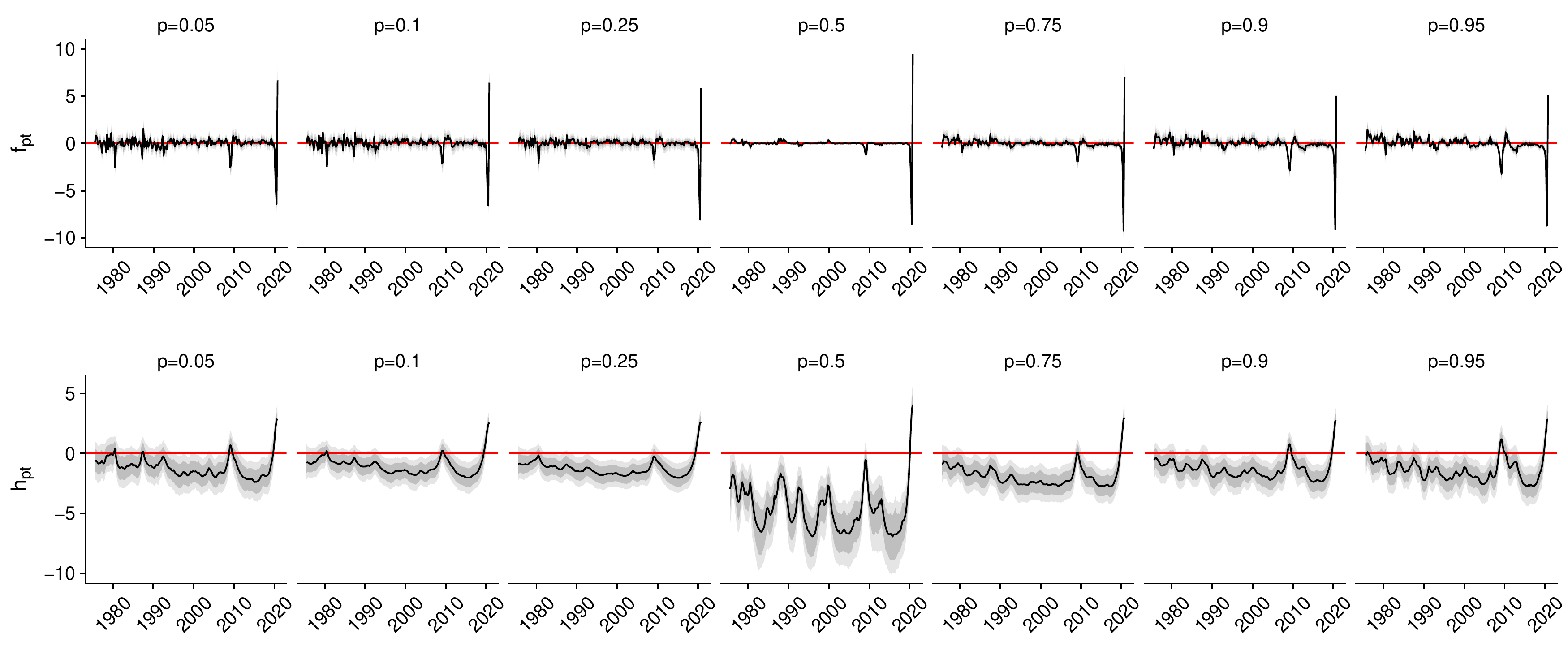}
    \caption{Estimates of the factors and associated log-volatility per quantile, HSP prior.}
    \label{fig:factors}
\end{figure}

The heterogeneity across quantiles in the role of the volatility factor is further supported by Figure \ref{fig:factors}, which reports estimates of the factors (upper panels) and associated log-volatility per quantile (lower panels).  In the upper panel, we observe that especially in the tails the factor moves sharply during global events such as the global financial crisis and the Covid-19 pandemic. To a somewhat smaller extent the results also suggest declines in the beginning of the 1990s and the early 2000s. When we focus attention on the 50 percent quantile we find strikingly different results. In the center of the distribution, the factor is small and very close to zero throughout the sample. During the pandemic we find a strong pronounced decrease in 2020:Q2, which was triggered by an unprecedented downturn in real activity globally but also a strong increase in 2020:Q3 (which was accompanied with sharply increasing GDP growth rates throughout all our countries).

Turning to the evolution of the log-volatilities in the lower panel generally yields consistent insights with the findings discussed for the level of the factor. The log-volatility spikes during recessions (i.e., in the early 1990s, 2008-2009, and 2020), and for $p=0.5$ the level of the log-volatility is much smaller than for the other quantiles but then  exceeds the increases in volatility observed for the other quantiles of the distribution. This finding also sheds light on why the amount of variation explained through the factor for most countries is lowest but still sizable in the 50 percent quantile. In most periods, the volatility factor is small (around $-5$ to $-10$ on the log-scale) but then during the pandemic it rapidly increases and reaches values of around 5 on the log-scale. This suggests that in tranquil periods, the factor only explains little variation in the shocks but in recessions (or turbulent times) this share increases appreciably and approaches 1.

%Here in particular the results at the 50 percent quantile are rather different from those at the other quantiles.  \todd{Particularly in the tail quantiles, the factor and its volatility move most sharply around the Great Recession and in 2020.}
\subsection{Generalized impulse responses to a global business cycle shock}
The discussion on the qualitative and quantitative properties of the estimated factor provides evidence that it can be interpreted as a global business cycle shock since, depending on the quantile adopted, it closely tracks events such as global recessions. Following \citet{SW2005JEEA}, we now consider how changes to the factor, labeled factor shocks, impact GDP growth across countries and quantiles.

  \begin{figure}[t]
    \centering
    \includegraphics[width=0.85\textwidth]{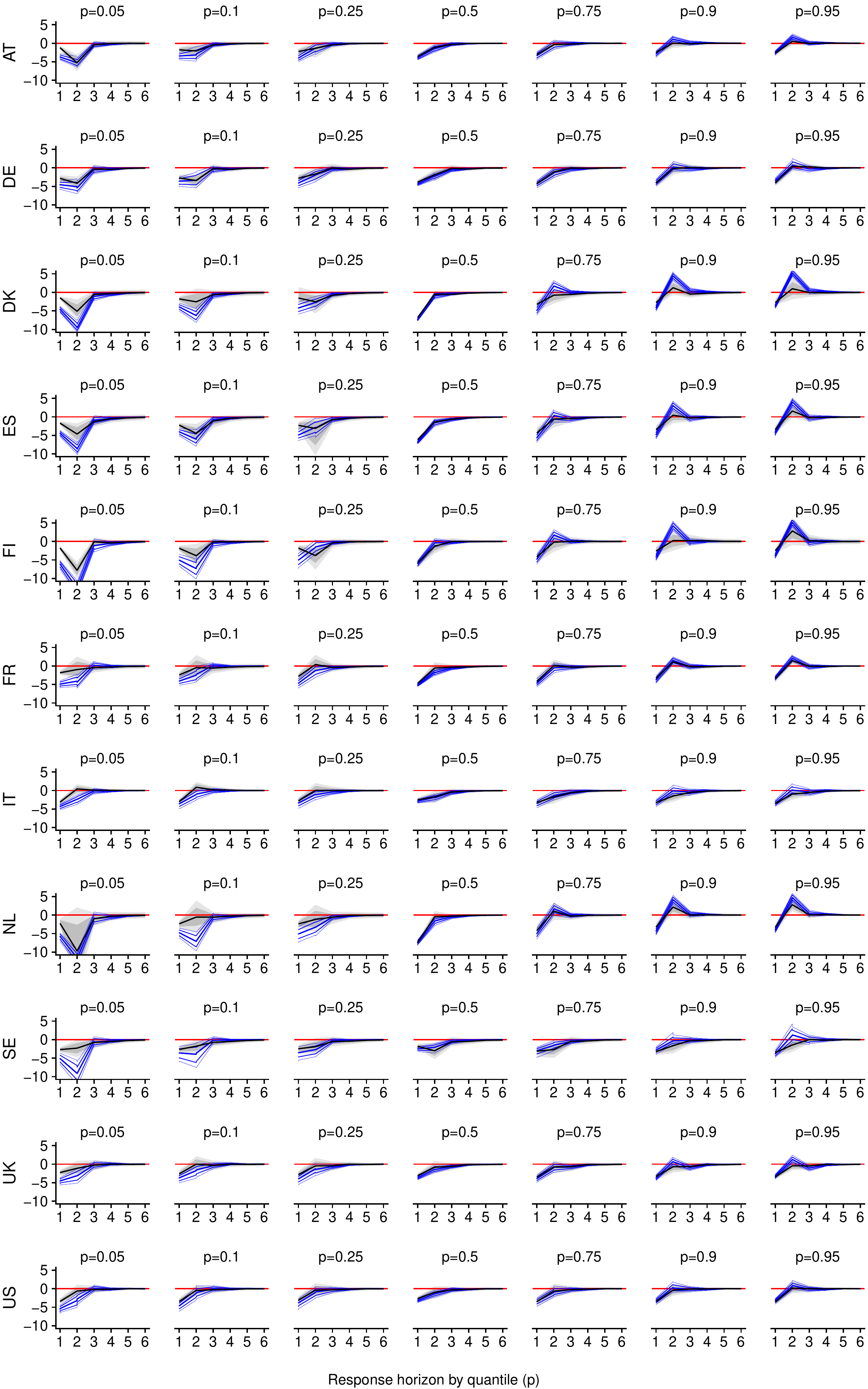}
    \caption{Generalized impulse response functions to a unit shock in $f_{pt}$ across countries and quantiles (average over time). \texttt{CISS-CC} with estimated $\omega$ in grey (16th, 84th percentiles), parametric \texttt{CISS-CC} with $\omega=0$ in blue (16th and 84th percentiles), HSP prior. The black and dark blue lines refer to the posterior median, respectively.}
    \label{fig:IRF-shock}
\end{figure}

The posterior quantiles of the generalized impulse response functions (GIRFs) for a common factor shock as estimated with the CISS-CC-HSP model are reported in Figure \ref{fig:IRF-shock}. This figure includes the GIRFs for our model with $\omega_{ip}$ estimated (gray shaded areas) and for $\omega_{ip}=0$ (solid blue lines).

 A first interesting finding is that, for all countries, a factor shock has different effects in the left tail than the right tail. In both tails, growth is negatively affected,  confirming that higher volatility/uncertainty is detrimental for growth, but the size of the effect (and persistence of the negative effect) is much larger in the left tail. Moreover, notice that for the right tail we also observe an overshoot in real activity in response to an adverse business cycle shock.
 
 Second, when we consider  the left tail, the BART piece with an estimated weight tends to mitigate the effects of the shock. This is most likely driven by the fact that, if we rule out non-linearities, there is more to be explained through the factor model and this might translate into factor dynamics which not only pick up business cycle shocks but also soak up information left in the error term potentially arising from ignoring non-linear dynamics between GDP growth and the CISS.
 
 A third striking pattern is the pronounced degree of cross-country heterogeneity in the 5 percent quantile (and, to a somewhat lesser extent, in the 10 percent quantile). When our focus is on the left tail, we observe that France, Italy, the UK, and Spain exhibit sharp declines in GDP growth.  Once we consider higher quantiles the GIRFs become much more similar across countries. For instance,  we find only modest differences if we focus on $p=0.5$.
 
 To sum up, we find that the countries in our sample display pronounced reactions to an international business cycle shock. These reactions differ not only across quantiles but also across countries.

 %It is worth stressing that, especially for $p=0.05$, the declines in GDP growth are much more pronounced if the combination weight is set to zero (i.e.,a linear model is used). This is most likely driven by the fact that if we rule out non-linearities there is more to be explained through the factor model and this might translate into factor dynamics which do not only pick up business cycle shocks but also soak up information left in the error term potentially arising from ignoring non-linear dynamics between GDP growth and the CISS.
 \clearpage
 \subsection{Generalized impulse responses to a US financial conditions shock}
 The previous sub-section emphasized that our latent factor can be interpreted as a global business cycle shock. In this sub-section we will instead focus attention on the international effects of a shock to US financial conditions and whether the real effects of such a shock differ from the ones arising from changes in  $f_{pt}$.
 
\begin{figure}[h!]
    \includegraphics[width=\textwidth]{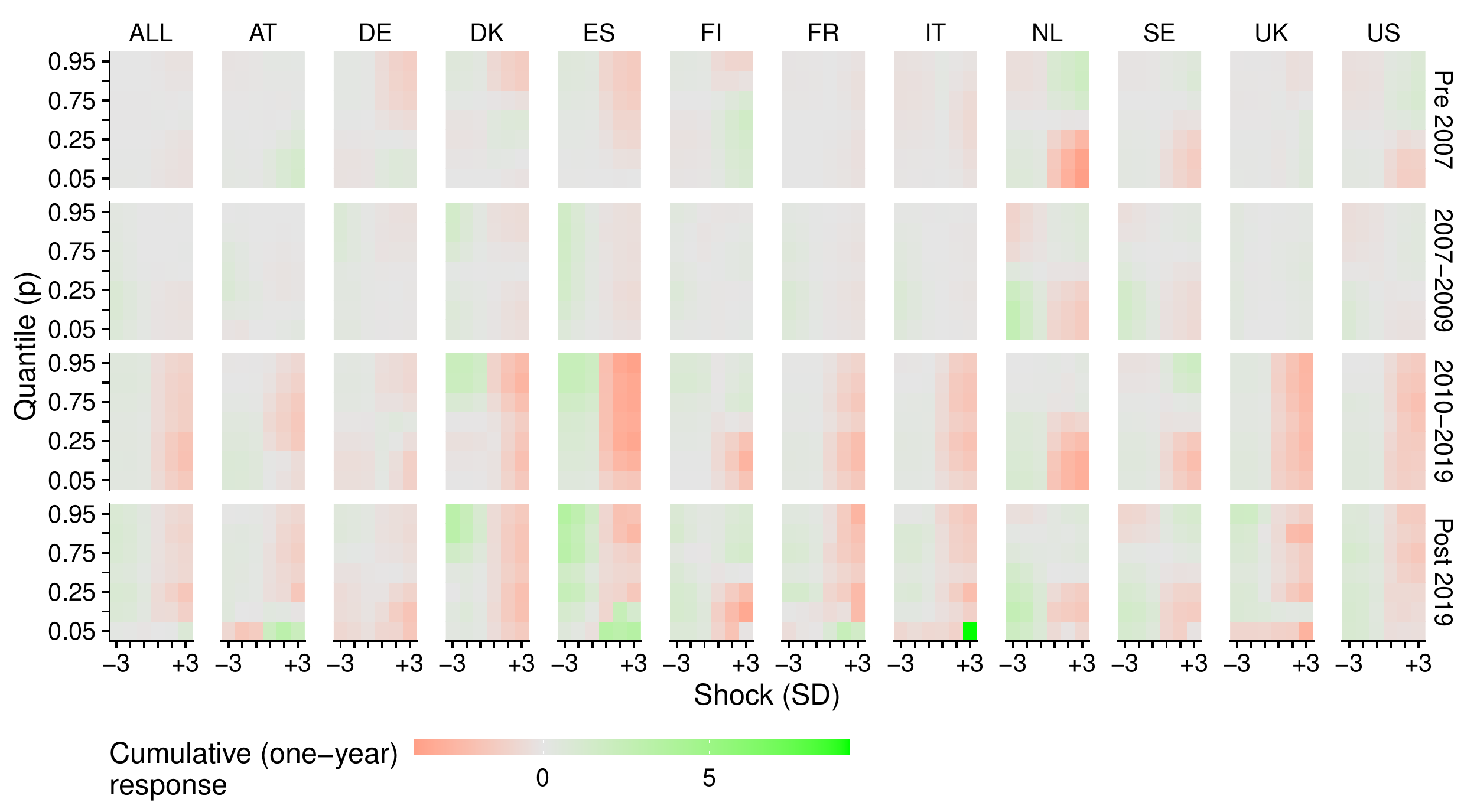}
    \caption{Cumulative (one-year ahead) generalized impulse response functions of GDP by quantile (p) to a $\{-3,-2,-1,1,2,3\}$ in-sample standard deviation US financial conditions shock. Model: \texttt{CISS-CC}, $\bm{\Lambda}\neq0$ and estimated $\bm{\omega}$, HSP prior. Difference between conditional and unconditional forecast across countries (grouped average over time by indicated period).}\label{fig: GIRF_lambdaon}
\end{figure}
\begin{figure}[h!]
    \includegraphics[width=\textwidth]{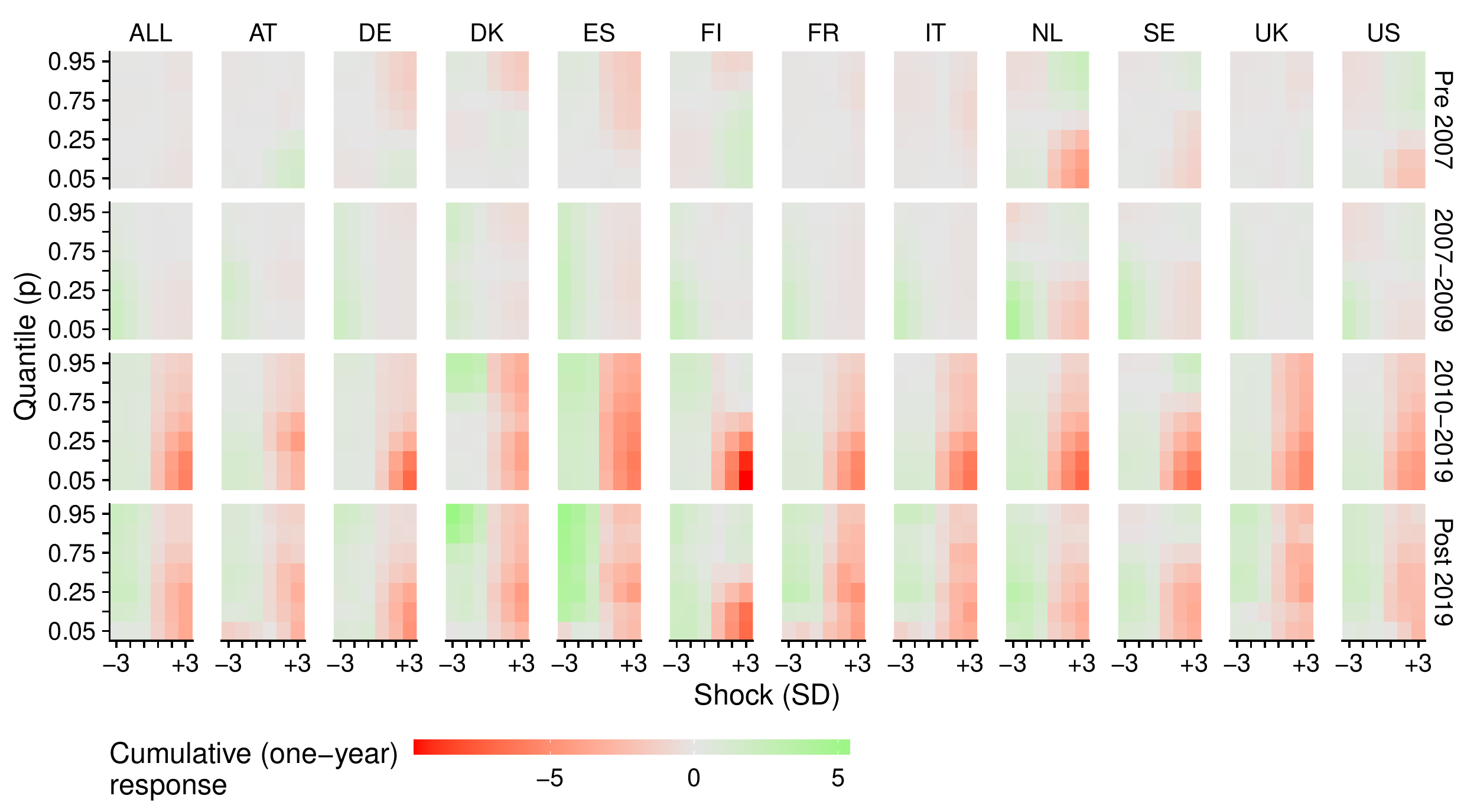}
    \caption{Cumulative (one-year ahead) generalized impulse response functions of GDP by quantile (p) to a $\{-3,-2,-1,1,2,3\}$ in-sample standard deviation US financial conditions shock. Model: \texttt{CISS-CC}, $\bm{\Lambda}=0$ and estimated $\bm{\omega}$, HSP prior. Difference between conditional and unconditional forecast across countries (grouped average over time by indicated period).}\label{fig: GIRF_lambdaoff}
\end{figure}

Figures \ref{fig: GIRF_lambdaon} and \ref{fig: GIRF_lambdaoff} report the posterior median of the  cumulative (one-year ahead) generalized impulse response functions of GDP by quantile ($p$) to a $-3$, $-2$, $-1$,  $1$, $2$ and $3$ in-sample standard deviation US financial conditions shock, based on the CISS-CC-HSP model with estimated weight $\omega$ and either with (Figure \ref{fig: GIRF_lambdaon}) or without (Figure \ref{fig: GIRF_lambdaoff}) the common factor in the model's innovation component. Recall that the CISS is defined so that higher values represent tighter financial conditions; a positive shock may be expected to reduce GDP growth. 

As the model is non-linear, the sign and size of the shocks can matter to determine the effects  (i.e., contrary to the linear case, the effects are not proportional to the size of the shock, or symmetric). Hence, we have recomputed the model for various sub-samples to analyze how different global business cycle conditions impact the estimates of the GIRFs.

Comparing the two figures shows that the factor volatility has little effect on the results, which is not surprising as it should not affect much the point estimates of the GIRFs (rather their precision). In both cases, a financial shock  in the US spills over to other countries. There is asymmetry in the sense that a positive shock affects the growth quantiles, whereas a negative shock’s effects are not as sharp. Moreover, the effects and asymmetry are sharper in the 2010-2019 period than earlier.

\begin{figure}[h!]
    \includegraphics[width=\textwidth]{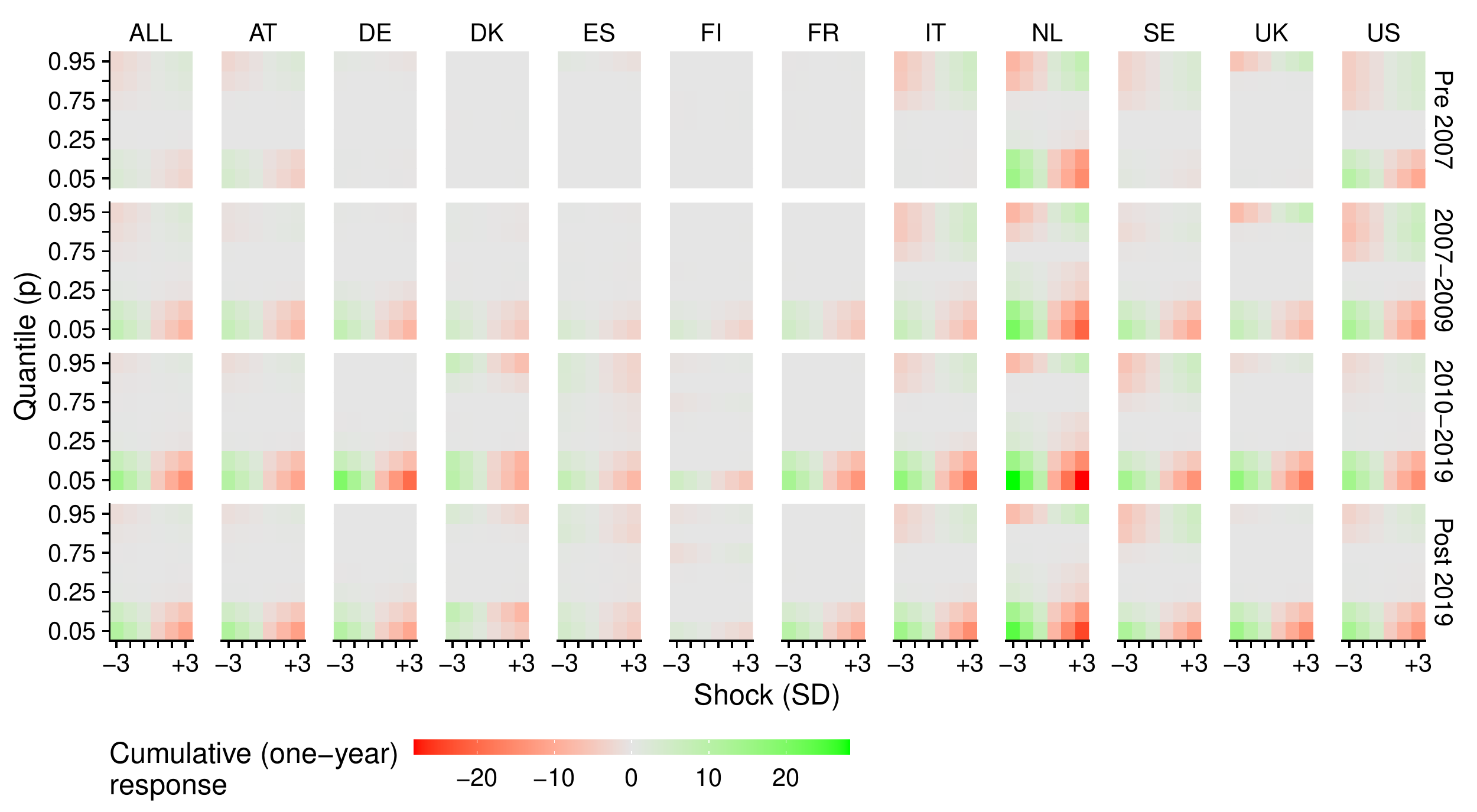}
    \caption{Cumulative (one-year ahead) generalized impulse response functions of GDP by quantile (p) to a $\{-3,-2,-1,1,2,3\}$ in-sample standard deviation US financial conditions shock. Model: \texttt{CISS-CC}, $\bm{\Lambda}=0$ and $\bm{\omega}=\bm{0}$, HSP prior. Difference between conditional and unconditional forecast across countries (grouped average over time by indicated period).}\label{fig: girf_ciss_weight0}
\end{figure}

\begin{figure}[h!]
    \includegraphics[width=\textwidth]{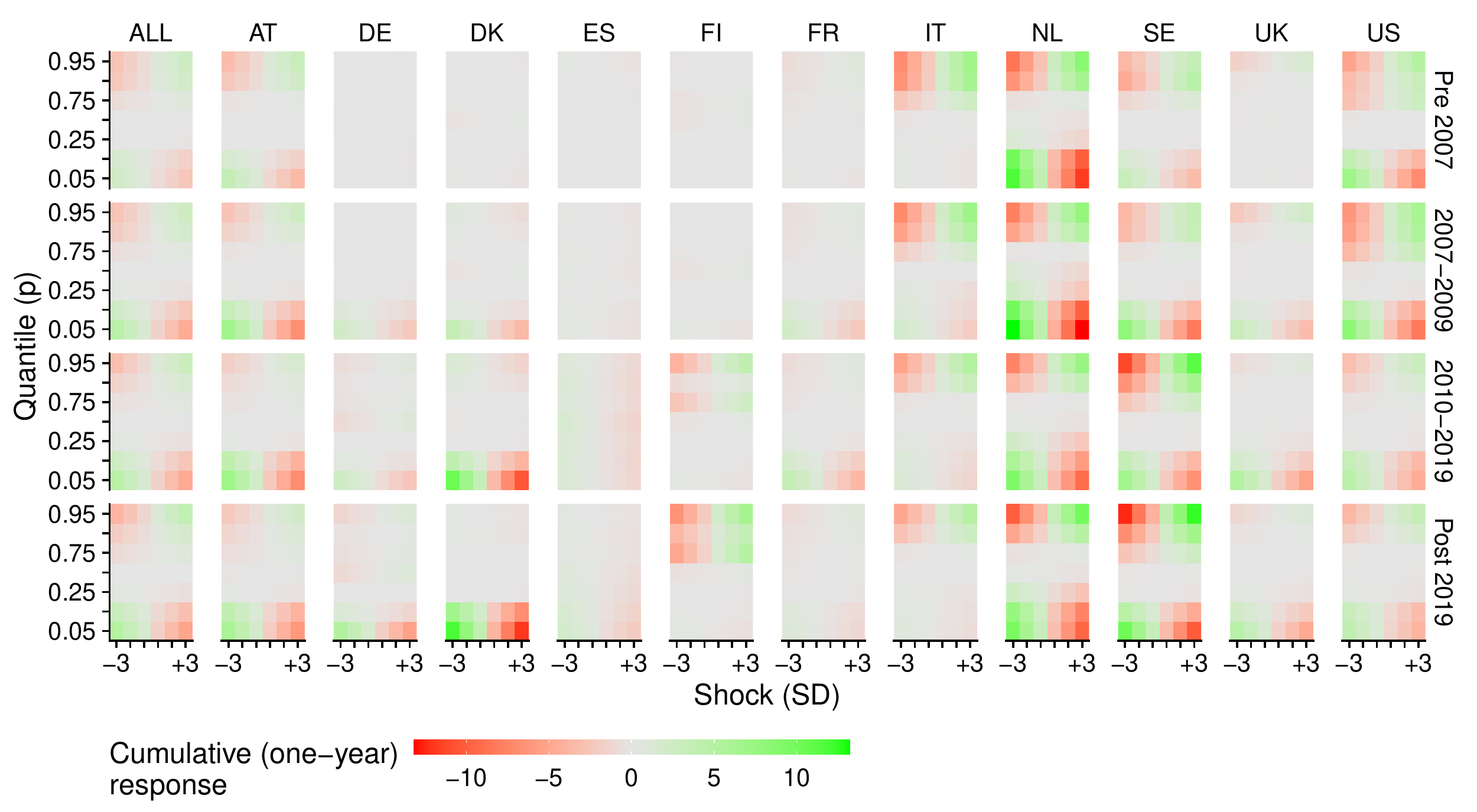}
    \caption{Cumulative (one-year ahead) generalized impulse response functions of GDP by quantile (p) to a $\{-3,-2,-1,1,2,3\}$ in-sample standard deviation US financial conditions shock. Model: \texttt{CISS-CC}, $\bm{\Lambda}\neq0$ and $\bm{\omega}=\bm{0}$, HSP prior. Difference between conditional and unconditional forecast across countries (grouped average over time by indicated period).}\label{fig: girf_ciss_weightneq0}
\end{figure}

Analyzing cross-country differences provides additional interesting insights.  For some countries (DE, FR, DK, and ES), we find substantial time variation in the GIRFs. Prior to 2010, the corresponding heatmaps feature a great deal of gray colored cells, implying no reactions at all. After the global financial crisis, US-CISS shocks have pronounced effects for these countries that are mostly located in the left tail of the distribution of GDP growth. For other countries (IT,  UK, US, and NL), we find less evidence in favor of time-variation in the GIRFs. In these countries, positive (negative) shocks to the US-CISS have negative (positive) effects on GDP growth for $p \in \{0.05, 0.1\}$. Notice, however, that when we consider $p \in \{0.9, 0.95\}$, the effect of a CISS shock seems to reverse sign; a positive shock to the CISS has positive effects on GDP growth if it is already historically high and a negative shock triggers a decline in GDP growth.  %\textbf{Flo: this is extremely interesting since it essentially implies that if GDP growth is very high, improved financial conditions in period $t$ seem to trigger a decrease in $t+1$, deleveraging?}

Figures \ref{fig: girf_ciss_weight0} and \ref{fig: girf_ciss_weightneq0} are similar to Figures  \ref{fig: GIRF_lambdaon} and \ref{fig: GIRF_lambdaoff}  but now the weight $\omega$ is set to zero, so that the quantile part is linear. Here, too, the common (volatility) factor component does not seem to have an obvious effect on the results. Instead, the effects of financial conditions are now more linear; positive and negative shocks to financial conditions in the US have similar effects, of opposite sign.  In addition, the pattern is clearer in the data since 2007 than before.  The effects also look smaller and more stable over time than in the non-linear quantile specification, and in general they are more marked at the lower quantiles.  These marked differences in the empirical findings highlight the importance of allowing for non-linearities also in the context of quantile regressions.

% ----------------------------------------------------------------------------------------------

%Generalized impulse responses to a shock of the common factor. Results are based on the \texttt{CISS-CC} (assuming financial conditions to remain constant). We identify the scale of the factors by normalizing each draw (in the algorithm) to have zero mean and unit variance. Indicated posterior credible sets are 16/84 and 5/95 percent (for \texttt{CISS-CC} with estimated $\omega$ in grey shades). The blue lines are the corresponding posterior credible set for the model with $\omega = 0$.

%Conditional forecasts are computed from specifications using $h=1,2,\hdots,24$. We use the in-sample unconditional standard deviation (SD) $\sigma_{xi}$ in country $i$ to simulate shocks for $s\in \{-3,-2,-1,1,2,3\}$ SDs. For unconditional forecasts, we simply use the $T$th observation, e.g., $x_{iT}$. The assumption is that the CISS measure for the US increases/decreases by $3$ in-sample standard deviations. Conditional forecasts are computed such that $\tilde{x}_{\text{US},T} = x_{\text{US},T} + s_l\sigma_{x,\text{US}}$ for $l = 1,2$. Impulses over time are in the Appendix.

\section{Conclusions}
In this paper we propose a non-parametric quantile panel regression model which assumes that the conditional mean is a convex combination of a linear and an unknown non-linear function. We learn the unknown functions using BART, a successful tool closely related to random forests. To decide on how much weight the BART piece should receive in the $p^{th}$ quantile, we estimate it alongside the remaining model parameters.  This non-parametric feature enhances model flexibility, especially in the tails. Using cross-sectional information, in addition, enables us to improve predictive accuracy. This is achieved by proposing a novel pooling prior as well as introducing cross-country information directly.  To carry out estimation and inference we design a scalable MCMC algorithm and apply the model to investigate "growth at risk" using an international panel of 11 countries.

In terms of empirical results, our proposed models commonly improve on the benchmark single country linear quantile model in recursive growth forecast comparisons, more so in the tails than near the center of the distribution and in particular when estimating the weight $\omega$, in turn allowing for both linear and BART pieces in the model. The estimated combination weight is smaller (i.e., the model is more linear) for the 25 and 75 percent quantiles than in the tails, i.e., for the 5 and 95 percent quantiles. Moreover, some form of international information definitely pays off (either via the new pooling prior, or by outright including non-domestic series). The effects of the common (volatility) factor are also relevant, as it seems to explain a large fraction of the forecast error variance in most countries, in particular in the tails. A shock to this factor, which can be interpreted as an uncertainty shock, has different (stronger negative) effects in the left tail than the right tail of the growth distribution.   In the left tail, the BART piece with an estimated weight tends to mitigate a bit the effects of the shock. Finally, a financial shock in the US spills over to other countries. There is asymmetry in the responses in the sense that a positive shock affects the growth quantiles, whereas a negative shock’s effects are not as sharp.  Moreover, the effects and asymmetry are sharper in the 2010-20 period than earlier.  The responses are instead much more proportional and symmetric in the linear model, highlighting the importance of allowing for non-linearities in the specification of quantile regressions.

\small{\setstretch{0.85}
\addcontentsline{toc}{section}{References}
\bibliographystyle{frbcle.bst}
\bibliography{lit}}\normalsize\clearpage

@article{figueres2020vulnerable,
  title={Vulnerable growth in the euro area: Measuring the financial conditions},
  author={Figueres, Juan Manuel and Jaroci{\'n}ski, Marek},
  journal={Economics Letters},
  volume={191},
  pages={109126},
  year={2020},
  publisher={Elsevier}
}

@article{SW2005JEEA,
    author = {Stock, James H. and Watson, Mark W.},
    title = {Understanding Changes in International Business Cycle Dynamics},
    journal = {Journal of the European Economic Association},
    volume = {3},
    pages = {968-1006},
    year = {2005},
    doi = {10.1162/1542476054729446},
}

@article{chkmp2021tail,
    author = {Todd E. Clark and Florian Huber and Gary Koop and Massimiliano Marcellino and Michael Pfarrhofer},
    title = {Tail Forecasting with Multivariate Bayesian Additive Regression Trees},
    journal = {FRB of Cleveland Working Paper},
    year = {2021},
    volume = {21-08},
    doi = {10.2139/ssrn.3809866}
}

@article{delle2020modeling,
  title={Modeling and forecasting macroeconomic downside risk},
  author={Delle Monache, Davide and De Polis, Andrea and Petrella, Ivan},
  journal={CEPR Discussion Paper Series},
  number ={15109},
  year={2020}
}

@article{kozumi2011gibbs,
  title={Gibbs sampling methods for Bayesian quantile regression},
  author={Kozumi, Hideo and Kobayashi, Genya},
  journal={Journal of statistical computation and simulation},
  volume={81},
  number={11},
  pages={1565--1578},
  year={2011},
  publisher={Taylor \& Francis}
}

@article{clark2021tail,
  title={Tail Forecasting with Multivariate Bayesian Additive Regression Trees},
  author={Clark, {Todd E} and Huber, Florian and Koop, Gary and Marcellino, Massimiliano and Pfarrhofer, Michael},
  year={2021},
  journal={FRB of Cleveland Working Paper},
  volume={21-08}
}

@article{makalic2015simple,
  title={A simple sampler for the horseshoe estimator},
  author={Makalic, Enes and Schmidt, Daniel F.},
  journal={IEEE Signal Processing Letters},
  volume={23},
  number={1},
  pages={179--182},
  year={2015},
  doi = {10.1109/LSP.2015.2503725},
  publisher={IEEE}
}

@article{yu2001bayesian,
  title={Bayesian quantile regression},
  author={Yu, Keming and Moyeed, Rana A},
  journal={Statistics \& Probability Letters},
  volume={54},
  number={4},
  pages={437--447},
  year={2001},
  publisher={Elsevier}
}

@article{kastner2014ancillarity,
  title={Ancillarity-sufficiency interweaving strategy ({ASIS}) for boosting {MCMC} estimation of stochastic volatility models},
  author={Kastner, Gregor and Fr{\"u}hwirth-Schnatter, Sylvia},
  journal={Computational Statistics \& Data Analysis},
  volume={76},
  pages={408--423},
  year={2014},
  doi={10.1016/j.csda.2013.01.002},
  publisher={Elsevier}
}

@article{adrian2019vulnerable,
  title={Vulnerable growth},
  author={Adrian, Tobias and Boyarchenko, Nina and Giannone, Domenico},
  journal={American Economic Review},
  volume={109},
  number={4},
  pages={1263--89},
  year={2019},
  doi = {10.1257/aer.20161923}
}

@article{adrian2018term,
  title={The term structure of growth-at-risk},
  author={Adrian, Tobias and Grinberg, Federico and Liang, Nellie and Malik, Sheheryar},
  year={2018},
  journal={IMF Working Paper},
  volume={18/180},
  doi = {10.5089/9781484372364.001}
}

@article{cook2019assessing,
  title={Assessing macroeconomic tail risks in a data-rich environment},
  author={Cook, Thomas and Doh, Taeyoung},
  journal={Federal Reserve Bank of Kansas City Research Working Paper},
  volume={19-12},
  year={2019},
  doi = {10.18651/RWP2019-12}
}

@article{denicolo2017forecasting,
  title={Forecasting tail risks},
  author={De Nicol{\`o}, Gianni and Lucchetta, Marcella},
  journal={Journal of Applied Econometrics},
  volume={32},
  number={1},
  pages={159--170},
  year={2017},
  doi = {10.1002/jae.2509},
  publisher={Wiley Online Library}
}

@techreport{Ferrara2019,
  title={Real-time high frequency monitoring of growth-at-risk},
  author={Ferrara, Laurent and Mogliani, M. and Sahuc, J.G.},
  year={2019}
}

@article{gaglianone2012constructing,
  title={Constructing density forecasts from quantile regressions},
  author={Gaglianone, Wagner Piazza and Lima, Luiz Renato},
  journal={Journal of Money, Credit and Banking},
  volume={44},
  number={8},
  pages={1589--1607},
  year={2012},
  doi = {10.1111/j.1538-4616.2012.00545.x},
  publisher={Wiley Online Library}
}

@article{galbraith2019asymmetry,
  title={Asymmetry in unemployment rate forecast errors},
  author={Galbraith, John W. and {van Norden}, Simon},
  journal={International Journal of Forecasting},
  volume={35},
  number={4},
  pages={1613--1626},
  year={2019},
  doi = {10.1016/j.ijforecast.2018.11.006},
  publisher={Elsevier}
}

@article{ghysels2018quantile,
  title={Quantile-based inflation risk models},
  author={Ghysels, Eric and Iania, Leonardo and Striaukas, Jonas},
  journal={National Bank of Belgium Research Working Paper},
  volume={349},
  year={2018},
  url={http://hdl.handle.net/10419/207729}
}

@article{giglio2016systemic,
  title={Systemic risk and the macroeconomy: An empirical evaluation},
  author={Giglio, Stefano and Kelly, Bryan and Pruitt, Seth},
  journal={Journal of Financial Economics},
  volume={119},
  number={3},
  pages={457--471},
  year={2016},
  doi = {10.1016/j.jfineco.2016.01.010},
  publisher={Elsevier}
}

@article{gonzalez2019growth,
  title = {Growth in stress},
  journal = {International Journal of Forecasting},
  volume = {35},
  number = {3},
  pages = {948-966},
  year = {2019},
  author = {Gloria Gonz\'alez-Rivera and Javier Maldonado and Esther Ruiz},
  doi = {10.1016/j.ijforecast.2019.04.006}
}

@article{fhkp2021,
  title={Approximate Bayesian inference and forecasting in huge-dimensional multi-country VARs},
  author={Feldkircher, Martin and Huber, Florian and Koop, Gary and Pfarrhofer, M.},
  year={2021},
  journal={arXiv},
  volume={2103.04944},
  url={https://arxiv.org/abs/2103.04944}
}

@article{kiley2018unemployment,
  title={Unemployment risk},
  author={Kiley, Michael T.},
  year={2018},
  journal={Board of Governors of the Federal Reserve System Finance and Economics Discussion Series},
  volume={2018-067},
  doi = {0.17016/FEDS.2018.067}
}

@article{korobilis2017quantile,
  title={Quantile regression forecasts of inflation under model uncertainty},
  author={Korobilis, Dimitris},
  journal={International Journal of Forecasting},
  volume={33},
  number={1},
  pages={11--20},
  year={2017},
  doi = {10.1016/j.ijforecast.2016.07.005},
  publisher={Elsevier}
}

@article{manzan2015forecasting,
  title={Forecasting the distribution of economic variables in a data-rich environment},
  author={Manzan, Sebastiano},
  journal={Journal of Business and Economic Statistics},
  volume={33},
  number={1},
  pages={144--164},
  year={2015},
  doi = {10.1080/07350015.2014.937436},
  publisher={Taylor \& Francis}
}

@article{manzan2013macroeconomic,
  title={Are macroeconomic variables useful for forecasting the distribution of {US} inflation?},
  author={Manzan, Sebastiano and Zerom, Dawit},
  journal={International Journal of Forecasting},
  volume={29},
  number={3},
  pages={469--478},
  year={2013},
  doi = {10.1016/j.ijforecast.2013.01.005},
  publisher={Elsevier}
}

@article{manzan2015asymmetric,
  title={Asymmetric quantile persistence and predictability: the case of {US} inflation},
  author={Manzan, Sebastiano and Zerom, Dawit},
  journal={Oxford Bulletin of Economics and Statistics},
  volume={77},
  number={2},
  pages={297--318},
  year={2015},
  doi = {10.1111/obes.12065},
  publisher={Wiley Online Library}
}

@article{mazzi2019nowcasting,
  title={Nowcasting Euro area {GDP} growth using quantile regression},
  author={Mitchell, James and Poon, Aubrey and Mazzi, Gian Luigi},
  year={forthcoming},
  journal={Advances in Econometrics}
}

@article{pfarrhofer2021,
  title={Tail forecasts of inflation using time-varying parameter quantile regressions},
  author={Pfarrhofer, Michael},
  year={2021},
  journal={arXiv},
  volume={2103.03632},
  url={https://arxiv.org/abs/2103.03632}
}

@article{korobilis2021,
  title={The time-varying evolution of inflation risks},
  author={Korobilis, Dimitris and Landau, Bettina and Musso, Alberto and Phella, Anthoulla},
  year={2021},
  journal={manuscript}
}

@article{plagborg2020growth,
  title={When is growth at risk?},
  author={Plagborg-M{\o}ller, Mikkel and Reichlin, Lucrezia and Ricco, Giovanni and Hasenzagl, Thomas},
  journal={Brookings Papers on Economic Activity},
  year={2020},
  pages={167 -- 229}
}

@article{reichlin2020financial,
  title={Financial variables as predictors of real growth vulnerability},
  author={Reichlin, Lucrezia and Ricco, Giovanni and Hasenzagl, Thomas},
  year={2020},
  journal={Deutsche Bundesbank Discussion Paper},
  volume={05/2020},
  url={http://hdl.handle.net/10419/214829}
}

@article{chipman1998bayesian,
  title={Bayesian {CART} model search},
  author={Chipman, Hugh A. and George, Edward I. and McCulloch, Robert E.},
  journal={Journal of the American Statistical Association},
  volume={93},
  number={443},
  pages={935--948},
  year={1998},
  doi = {10.2307/2669832},
  publisher={Taylor \& Francis}
}

@article{carriero2021addressing,
  title={Addressing {COVID}-19 outliers in {BVARs} with Stochastic Volatility},
  author={Carriero, Andrea and Clark, {Todd E.} and Marcellino, Massimiliano and Mertens, Elmar},
  journal={Federal Reserve Bank of Cleveland Working Papers},
  volume={21-02R},
  doi = {10.26509/frbc-wp-202102r},
  year={2021}
}

@article{chipman2010bart,
  title={{BART}: {Bayesian} additive regression trees},
  author={Chipman, Hugh A. and George, Edward I. and McCulloch, Robert E.},
  journal={The Annals of Applied Statistics},
  volume={4},
  number={1},
  pages={266--298},
  year={2010},
  doi={10.1214/09-AOAS285},
  publisher={Institute of Mathematical Statistics}
}

@article{gneiting2011comparing,
  title={Comparing density forecasts using threshold- and quantile-weighted scoring rules},
  author={Gneiting, Tilmann and Ranjan, Roopesh},
  journal={Journal of Business \& Economic Statistics},
  volume={29},
  number={3},
  pages={411--422},
  year={2011},
  doi = {10.1198/jbes.2010.08110},
  publisher={Taylor \& Francis}
}

@article{slicesampling,
  title={Slice sampling},
  author={Neal, Radford},
  journal={Annals of Statistics},
  volume={31},
  number={3},
  pages={705--767},
  year={2003}
}

@article{huber2020inference,
  title={Inference in {B}ayesian Additive Vector Autoregressive Tree models},
  author={Huber, Florian and Rossini, Luca},
  journal={arXiv},
  volume={2006.16333},
  year={2020},
  url={https://arxiv.org/abs/2006.16333}
}

@article{huber2020nowcasting,
  title={Nowcasting in a pandemic using non-parametric mixed frequency {VARs}},
  author={Huber, Florian and Koop, Gary and Onorante, Luca and Pfarrhofer, Michael and Schreiner, Josef},
  journal={Journal of Econometrics},
  volume={in-press},
  year={2020},
  doi = {10.1016/j.jeconom.2020.11.006},
  publisher={Elsevier}
}

@article{TaddyKottas2010JBES,
author = {Matthew Taddy and Athanasios Kottas},
title = {A Bayesian Nonparametric Approach to Inference for Quantile Regression},
journal = {Journal of Business and Economic Statistics},
volume = {28},
pages = {357-369},
year  = {2010},
doi = {10.1198/jbes.2009.07331},
}

@article{BaiCarrieroClarkMarcellino2020WP,
  title={Macroeconomic Forecasting in a Multi-country Context},
  author={Yu Bai and Carriero, Andrea and Clark, Todd and Marcellino,  Massimiliano},
  year={2020},
  journal={manuscript}
}

% ----------------------------------------------------------------------------------------------
\clearpage
\begin{center}
    \huge \textbf{Appendix}\\[1em]
    \LARGE \textbf{Further forecast results}
\end{center}

\begin{figure}[h]
    \includegraphics[width=\textwidth]{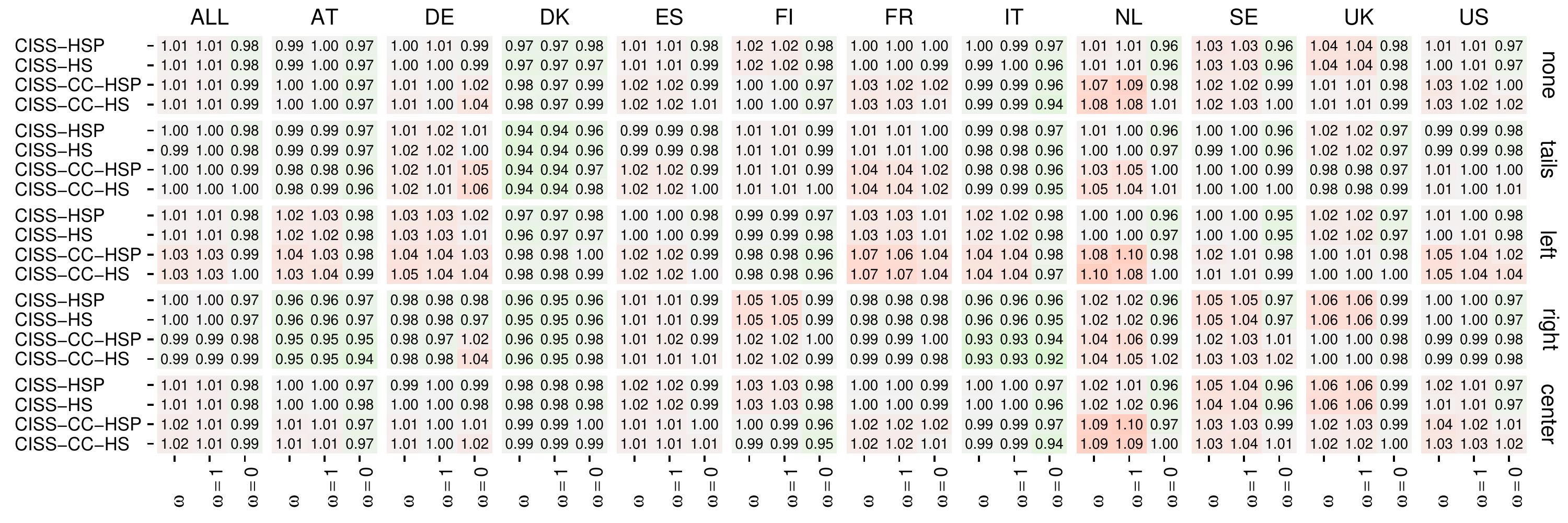}
    \caption{Relative quantile weighted cumulative ranked probability scores (CRPS) for $h=4$ and models with $\bm{\Lambda}\neq\bm{0}$. The results are benchmarked to ABG with $\bm{\Lambda}=\bm{0}$ and $\bm{\omega}=\bm{0}$. Lower ratios (shaded an green) indicate better performance (and vice versa, shaded in red).}
\end{figure}

\begin{figure}[h]
    \includegraphics[width=\textwidth]{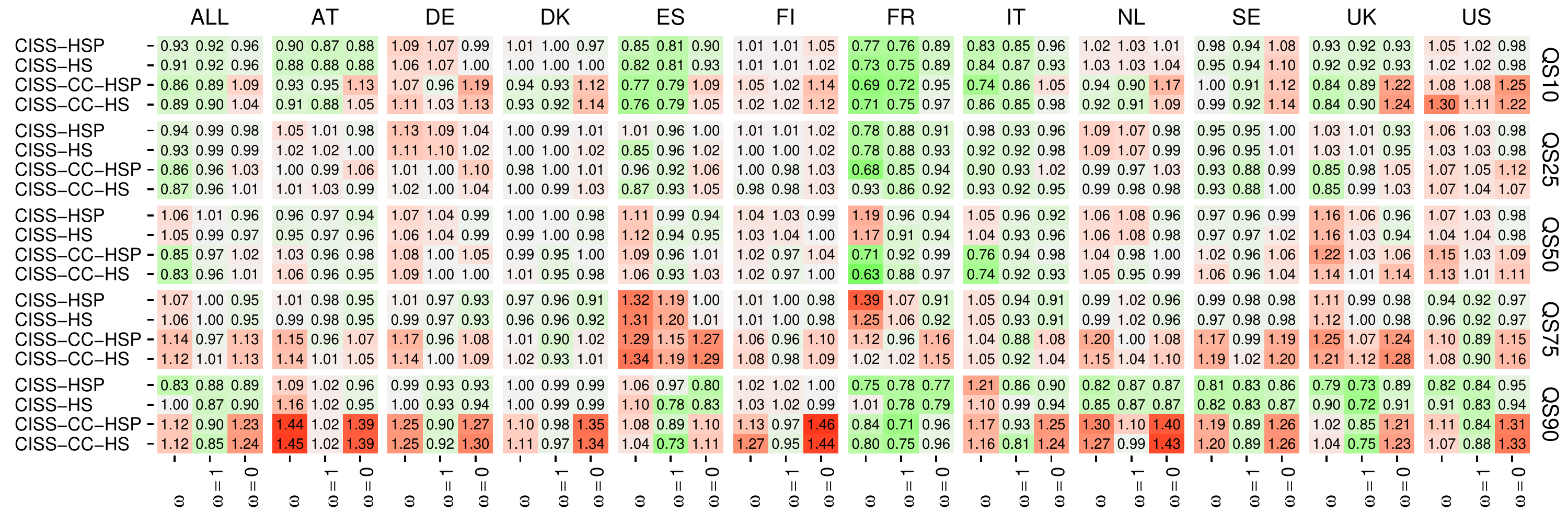}
    \caption{Relative quantile scores (QSs) for $h=1$ and models with $\bm{\Lambda}\neq\bm{0}$. The results are benchmarked to ABG with $\bm{\Lambda}=\bm{0}$ and $\bm{\omega}=\bm{0}$. Lower ratios (shaded an green) indicate better performance (and vice versa, shaded in red).}
\end{figure}

\begin{figure}[h]
    \includegraphics[width=\textwidth]{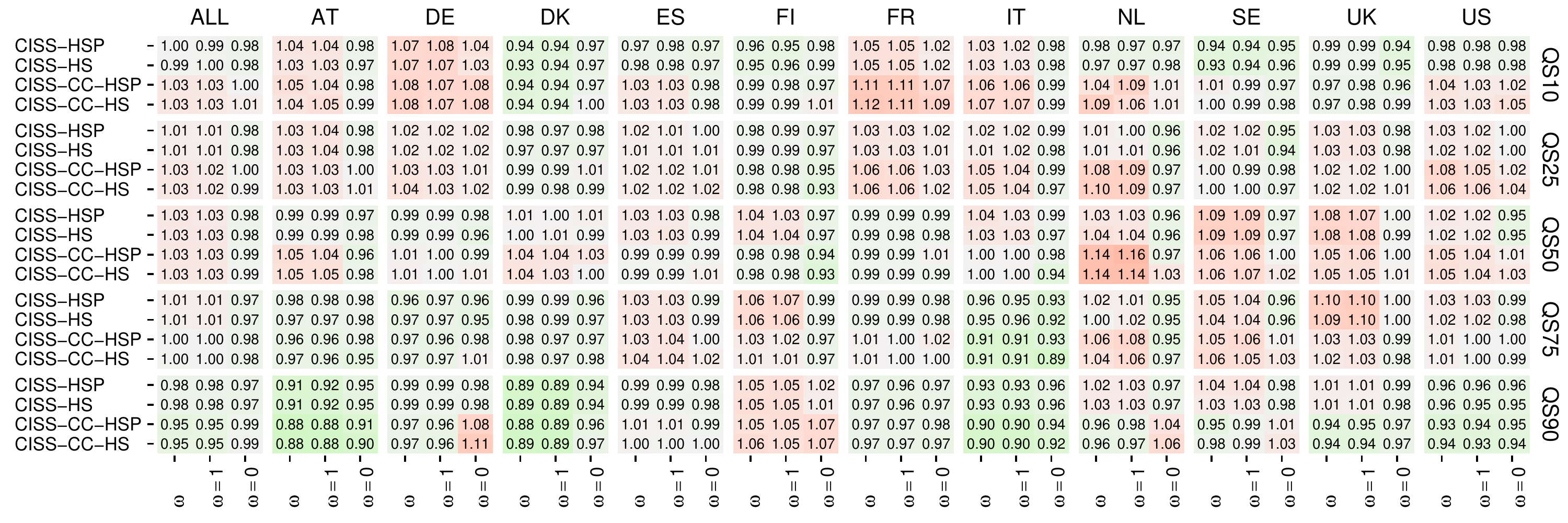}
    \caption{Relative quantile scores (QSs) for $h=4$ and models with $\bm{\Lambda}\neq\bm{0}$. The results are benchmarked to ABG with $\bm{\Lambda}=\bm{0}$ and $\bm{\omega}=\bm{0}$. Lower ratios (shaded an green) indicate better performance (and vice versa, shaded in red).}
\end{figure}

\end{document}